\documentclass[12pt]{article}
\usepackage{mathtools,amssymb,mathrsfs,ytableau,microtype,tikz,slashed}
\usepackage{physics}
\usepackage{color}
\usepackage[linktocpage]{hyperref}
\hypersetup{colorlinks=true,citecolor=blue,linkcolor=blue, urlcolor=blue, breaklinks=true}
\usepackage{cite}
\usepackage{moresize}
\usepackage{url}
\ytableausetup{centertableaux,boxsize=.5em}
\usepackage[numbers,sort]{natbib}

\makeatletter\@addtoreset{equation}{section}\makeatother

\newcommand{\B}{\mathcal{B}}
\newcommand{\C}{\mathbb{C}}
\newcommand{\Ha}{\mathcal{H}}
\newcommand{\St}{\mathcal{S}}
\newcommand{\M}{\mathcal{M}}

\def\Aut#1{{\text{Aut}_{\otimes}^{br}(#1)}}
\def\uuAut#1{\underline{\underline{\text{Aut}_{\otimes}^{br}(#1)}}}
\def\uuPic#1{\underline{\underline{\text{Pic}(#1)}}}

\renewcommand{\title}[1]{\vbox{\center\LARGE{#1}}\vspace{5mm}}
\renewcommand{\author}[1]{\vbox{\center\large#1}\vspace{5mm}}
\newcommand{\address}[1]{\vbox{\center\em#1}}

\begin{document}

\begin{titlepage}
\begin{center}
\vskip-10pt
{\Large \bf{On Topology of the
Moduli Space of
\\\vspace{0.5cm}Gapped Hamiltonians for Topological Phases
}}

\vspace{10mm}

 {Po-Shen Hsin$^a$\footnote{\href{mailto:phsin@phys.ucla.edu}{\tt phsin@physics.ucla.edu}}, 
 Zhenghan Wang,${}^{b,c}$\footnote{\href{mailto:zhenghwa@microsoft.com}{\tt zhenghwa@microsoft.com}}}

\vskip 7mm

\address{
${}^a$Mani L. Bhaumik Institute for Theoretical Physics,
475 Portola Plaza, Los Angeles, CA 90095, USA}

\address{
${}^b$ Station Q, Microsoft Research, Santa Barbara, California 93106-6105, USA
}

\address{
${}^c$ Department of Mathematics, University of California, Santa Barbara, California 93106, USA
}

\end{center}

\abstract{The moduli space of gapped Hamiltonians that are in the same topological phase is an intrinsic object that is associated to the topological order.  The topology of these moduli spaces is used recently in the construction of Floquet codes.  We propose a systematical program to study the topology of these moduli spaces.  In particular, we use effective field theory to study the cohomology classes of these spaces, which includes and generalizes the Berry phase.
We discuss several applications to studying phase transitions.
We show that nontrivial family of gapped systems with the same topological order can protect isolated phase transitions in the phase diagram, and we argue that the phase transitions are characterized by screening of topological defects.
We argue that family of gapped systems obey bulk-boundary correspondence.
We show that family of gapped systems in the bulk with the same topological order can rule out family of gapped systems on the boundary with topological order given by the topological boundary condition, constraining phase transitions on the boundary.
}

{\hypersetup{linkcolor=black}
\thispagestyle{empty}
}

\end{titlepage}

{\hypersetup{linkcolor=black}
\tableofcontents
\thispagestyle{empty}
}

\unitlength = .8mm

\setcounter{tocdepth}{3}

\medskip

\section{Introduction}

The theory of topological phases of matter has been an exciting development in both mathematics and physics for the last few decades with important applications to quantum computing (see {\it e.g.} 
\cite{Kitaev:1997wr,Freedman1998-eo,Freedman:2001,Dennis:2001nw,PhysRevA.71.022316,RevModPhys.80.1083,PhysRevLett.101.010501,Alicea_2011,PhysRevX.4.011036,qiu20}).
Toy models have played extremely important roles for the study of topological phases such as the celebrated toric code \cite{Kitaev:1997wr}. 
Topological phases can also embed in general quantum systems as the subsector of topological defects, or as regions in the phase diagram.
Topological defects  in general quantum systems play an important role in understanding the dynamics, such as the anomaly matching in the renormalization group flow that preserves the symmetry generated by the topological defects, deconfinement transition \cite{Komargodski:2020mxz} and incompatibility of the defect with a trivial phase \cite{Choi:2021kmx,Choi:2022zal}.
However, the constraints on the dynamics from general topological defects have not been fully understood.

In both mathematics and physics, the idea of a moduli space has become more and more fundamental. 
In this work, we will study the moduli space of the family of gapped systems that have the same low energy topological phase. 
Such space can be parameterized by continuous deformations in the microscopic model that preserve the low energy topological phase. 
Examples of such space are studied in {\it e.g.} \cite{Aasen:2022cdu,Kapustin:2020eby,Kapustin:2020mkl, Hsin:2020cgg}.  We will focus on topological phases described by topological quantum field theories (TQFT).
While it is not yet completely clear how to define such a topological space of gapped Hamiltonians mathematically, sufficient understanding is accumulated that we have a good working definition, especially in two spacial dimensions. 
 In the later part of the paper, we will generalize some of the discussions to higher dimensions that do not rely on a detailed  definition of the topological order in higher dimension.

As a working definition, we define a topological phase as a gapped Hamiltonian schema that realizes a given topological order, where the Hamiltonian has a finite energy gap separating the ground states from the excited states in the thermodynamic limit.
Two Hamiltonian schemas realize the same topological order if and only if they can be connected adiabatically by a path of gapped Hamiltonians without closing the spectral gap under suitable stabilization and coarse graining.  We define a topological order in spacial dimension $(n+1)$ as a unitary modular tensor $n$-category (UMTnC) with some extra data such as the chiral central charge in two spacial dimensions.  Such a UMTnC would lead to an $n$-extended $((n+1)+1)$-TQFT, for $n=1$ see \cite{turaev16}.  We do not have a mathematical definition of UMTnCs either except for $n=1$ and possibly $n=2$.  But this would suffice for our purpose as 2-spacial dimensions so $n=1$ is the main focus of this paper. For $n=1$, a UMTC $\B$ is called an anyon model and we will regard a genus $(\B,c)$ as a topological order\footnote{A genus is a pair of an anyon model $\B$ with a non-negative ration number $c$ such that $\frac{\sum_i\theta_i d_i^2}{D}=e^{2\pi ic/8}$.  The word genus comes from the case of abelian anyon models and lattice conformal field theories.  In this case, $c$ is the chiral central charge of the lattice CFT and the abelian anyon model of the lattice CFT is given by the genus of the lattice.}.  Then given a topological order $(\B,c)$, the moduli space of all gapped Hamiltonians that realize the genus $(\B,c)$ is the topological space that we are interested in.  

The topology of the space of gapped system gives rise to topological defects in the theory as discussed in \cite{Hsin:2020cgg}. When such space is described by some parameter $\lambda$ in the microscopic models that has the same low energy topological order, we can construct topological defects by varying the parameter over the spacetime. For instance, we can construct topological domain wall by varying the parameter on a circle that pierces the domain wall, whose value traces around a non-contractible one-cycle in the space of parameters. An example of such domain wall is discussed in \cite{Aasen:2022cdu} using a family of Hamiltonians that describe the untwisted $\mathbb{Z}_2$ gauge theory topological order, with the domain wall generating the electromagnetic duality symmetry.

Such a non-contractible path is also explored in the construction of Floquet code \cite{Hastings:2021ptn}, which is a subsystem code without static logical qubit, while periodic measurements generate an effective (``dynamical") logical qubit.

In this work, we will explore the generalizations to higher-dimensional cycles on the parameter space, and the properties of the corresponding topological defects. We can vary the parameter space over an $n$-dimensional cycles in spacetime with value tracing out a $k$-dimensional cycle on the parameter space with $k\leq n$, and this gives a codimension-$n$ topological defect that intersects with the $n$-dimensional cycle in spacetime. We will say such spacetime-dependent parameter traps the topological defect. When the defect is supported on a point, it can be a non-trivial phase times the identity operator, and this is an example of Berry phase discussed in \cite{Hsin:2020cgg}.\footnote{More generally, one can vary the parameter over an auxiliary bulk such that there is no bulk dependence, this produces the integer class Berry phase \cite{Kapustin:2020eby,Hsin:2020cgg}, and it is the analogue of Wess-Zumino term in nonlinear sigma model.}
If the defect is non-trivial, we have a non-trivial family of gapped systems. 

The topology of the space of gapped systems also probes quantum phase transitions. Examples of such application are discussed in \cite{Hsin:2020cgg}, where Berry phase on cycles in the phase diagram predicts the existence of diabolical phase transitions where the gap closes. In this work we will investigate phase transitions not protected by Berry phase, but by other information about the topology in the space of gapped systems.
Consider the phase diagram of quantum system that contain isolated phase transitions locally surrounded by topological phases with the same topological order. Then we can investigate the phase transitions using the family of gapped systems in the surrounding region in the phase diagram. For instance, consider an $n$-dimensional cycle in the parameter space of the gapped systems that traps a non-trivial topological defects, then in the full phase diagram, if the cycle can be trivialized, {\it i.e.} contractible in the full phase diagram, then in the process of contraction we must encounter a phase transition in the phase diagram. 
 Moreover, this leads to the prediction that at the phase transition the topological defect trapped by the cycle can be screened.\footnote{
An example of phase transition with such topological defect is free massless fermion, with the $\mathbb{Z}_2$ fermion parity symmetry gauged. This gives a free bosonic CFT that has emergent quantum $\mathbb{Z}_2$ symmetry generated by the Wilson line of the gauge field for the fermion parity symmetry; such Wilson line can end on the massless fermion field. An example is 1+1d Ising CFT, where the twisted Hilbert space with the $\mathbb{Z}_2$ symmetry line defect contains the fermion operator (see {\it e.g.} \cite{Chang:2018iay}).
 }

Bulk-boundary correspondence is an important tool that constrains the boundary dynamics from the bulk, and vice versa.
In the case that the bulk is an invertible phase with symmetry, the symmetry-preserving boundary has the corresponding 't Hooft anomaly for the symmetry.
When the bulk is a family of invertible gapped systems with non-trivial Berry phase, the boundary has phase transition protected by the Berry phase by the bulk-boundary correspondence discussed \cite{Hsin:2020cgg}, and it is called an anomaly in the space of couplings on the boundary in \cite{Cordova:2019jnf,Cordova:2019uob}. We will discuss the generalization of the bulk-boundary correspondence to the case that the bulk is a family of general gapped systems. We ask the following question: suppose the bulk is a family of gapped systems with the same bulk topological order that admits a topological boundary condition, can the boundary be a family of gapped systems with the same boundary topological order given by this topological boundary condition. We will show that depending on the family in the bulk, the answer is negative, {\it i.e.} the family of bulk gapped systems imply that the boundary has a phase transition at some parameter.

We will discuss various examples to illustrate these ideas, with the main focus on the space of gapped systems in 2+1d that have the same intrinsic topological order ${\cal B}$.
This space is conjectured to be the classifying space of the Picard 3-groupoid of ${\cal B}$, ${\cal M}_\B \simeq B\uuPic\B)$ \cite{Hsin:2020cgg, Aasen:2022cdu}. 
The fundamental group describes ordinary symmetry of ${\cal B}$, {\it i.e.} the braided autoequivalence of ${\cal B}$.
The second homotopy group describes Abelian anyons, and the third homotopy group describes topological point operators and torsional Berry phase.

The work is organized as follows. In Section \ref{sec:moduli}, we discuss the moduli space of gapped bosonic Hamiltonians.
In Section \ref{sec:topmoduli}, we discuss the homotopy type and cohomology groups of the moduli space. In Section \ref{sec:EFT}, we discuss methods to construct family of gapped systems that are non-invertible using effective field theory. In Section \ref{sec:examples}, we discuss examples of family of gapped systems, using both lattice models and field theories. In Section \ref{sec:phase transition} we discuss applications to the investigation of phase transitions in the phase diagrams.
In Section \ref{seC:bulk-boundary} we discuss the bulk-boundary correspondence, {\it i.e.} the implication of family of gapped systems in the bulk on the boundary. 

\section{Moduli spaces of gapped Hamiltonians}
\label{sec:moduli}
In this section, we mainly study spin/bosonic systems so if not explicitly stated, the Hamiltonians are local gapped on the Hilbert spaces of a collection of qudits.

\subsection{Topological order and topological phases}

Morally, we would define a topological order as the catch-all high category of all topological properties in a topological phase of matter.  By a topological phase of matter, we would mean a gapped Hamiltonian schema\footnote{By Hamiltonian schema, we mean a well-defined procedure to write down Hamiltonians from some celluations of a space manifold, potentially with some extra background information such as orientation, branching and framing.} whose ground states satisfy the topological stability conditions in \cite{bravyi10}.  These stability conditions can be interpreted as lattice versions of the disk and annulus axioms of extended TQFTs \cite{qiu20}.  Two Hamiltonian schemas represent the same topological phase of matter if they can be connected adiabatically by a path of gapped Hamiltonians without closing the spectral gap under suitable stabilization and coarse graining.  The precise meaning of stabiliziation and coarse graining is one subtlety in the definition of topological phases of matter and we will not elaborate on this issue further.  We would require also that two equivalent Hamiltonian schemas realize the same topological order in the sense below. 

There are at least two equivalent ways to define when a Hamiltonian schema realizes a given topological order.  We will explain the two special dimensions case here and the generalization to high dimensions is straightforward.  For two spacial dimensions, the chiral central charge $c$ in a genus $(\B,c)$ is a physical quantity that can be measured in the lab.  In the following we will ignore this important invariant by focusing only on the anyon model $\B$.  

By \cite{turaev16}, an anyon model $\B$ leads to a $1$-extended unitary $(2+1)$ TQFT called the Reshetikhin-Turaev type TQFTs.  Therefore, we will use the terms anyon model and $(2+1)$-TQFT interchangeably.  

Reshetikhin-Turaev type unitary TQFTs are defined as symmetric monoidal functors from the category of bordisms of space-time manifolds to the category of finitely dimensional Hilbert spaces.  
Then a Hamilonian schema realizes such a TQFT if the ground state Hilbert space of the Hamiltonian schema on any closed spacial surface is functorially the same as the TQFT Hilbert space.

Another sense of realization is focusing on elementary excitations in the plane.  If all elementary excitations of a Hamiltonian in the plane form an anyon model, then we would say that the Hamiltonian schema realizes the anyon model.

\subsection{Two fibrations}

Let $\B$ be an anyon model, and $\Ha_{\B}$ be the space of gapped Hamiltonians that realize $\B$.  There are serious mathematical issues for the definition of such spaces.  Besides some standard subtleties mentioned in \cite{Aasen:2022cdu}, another subtlety is related to renormalization group (RG) flow: if $\B$ is a double, then there is a non-empty subspace of $\Ha_{\B}$ consisting of local commuting projector Hamiltonians
\begin{equation}
    H=-\sum  \alpha_i P_i,\quad P_iP_j=P_jP_i,\quad P_i^2=P_i
\end{equation}
where $i$ labels the local projectors $P_i$ of finite support on the lattice (note $P_i$ themselves can depend on coupling parameters in a family of gapped systems), and $\alpha_i$ are couplings. 
For $\alpha_i>0$, the ground states are given by minimizing each local projector $-P_i$,  and they do not depend on the couplings $\alpha_i$. Similarly, for any $\alpha_i<0$, the ground states are given by minimizing $P_i$:
\begin{equation}
    \alpha_i>0:\quad P_i|\text{GS}\rangle=|\text{GS}\rangle;\quad 
    \alpha_i<0:\quad 
    P_i|\text{GS}\rangle=0~.
\end{equation}
Such ground states are the same as those of the Hamiltonian $H'=-\sum \text{sgn}(\alpha_i)P_i$; in particular, if we fix the sign of the couplings $\alpha_i$, the ground states wavefunctions do not depend on their detailed value.
Thus it is natural to conjecture that the RG flow provides a deformation retract of $\Ha_{\B}$ to this subspace.  Now suppose $\B$ has non-trivial central charge $mod\, 8$, it is widely believed there are no such commuting projector realizations.  Then where would the RG flow ends?  One possible answer is that the RG flows end at commuting projector Hamiltonians with infinite dimensional local Hilbert spaces\footnote{The second author thanks D. Ranard and X.-G. Wen for pointing out such a possibility.}.  

One potential way to realize this possibility is to start with the commuting projector Hamiltonians on bounding $3$-manifolds \cite{walkerwang},which can realize gapped phases with non-trivial central charge on the 2D boundary, 
and regard such model as a 2D model with infinite specifies labelled by the bulk direction. Denote the bulk coordinate by $z$ with boundary at $z=0$, the edges can be labelled by $e_\text{3D}=(e_\text{2D},z)$, where $e_\text{2D}$ are the edges on the plane of constant $z$. Then the effective local 2D Hilbert space is formally
\begin{equation}
    {\cal H}^\text{eff}_{e_\text{2D}}=\otimes_{z=0}^{L_z} {\cal H}_{(e_\text{2D},z)}~,
\end{equation}
where $L_z\rightarrow \infty$ is the size of the bulk direction.
The Hamiltonian of the 3D model can be re-phrased as a 2D  Hamiltonian for such local Hilbert space of dimension $L_z\rightarrow \infty$. Since the 3D Hamiltonian is a sum of local commuting projector, each term with finite support on the 3D lattice, the resulting 2D Hamiltonian is also a sum of local commuting projector, each term only depends on finitely many ``specifies" labelled by $z$.

There are two variations of $\Ha_{\B}$ that are both fibrations.  The first is instead of the space of gapped Hamiltonians, we could discuss using the space $\St_{\B}$  of gapped states that represent the anyon model $\B$.  It seems to be known to many people that these two spaces $\Ha_{\B}$ and $\St_{\B}$ should have the same homotopy type as there is a hypothetical fibration $F_{\B}\rightarrow \Ha_{\B} \rightarrow \St_{\B}$.  The fiber space $F_\B$ is the space of all gapped Hamiltonians that realize the same state as ground state, hence contractible.  

Closely related to our discussion is another conjectured fibration which comes from the space $E_{d=2}$ of all gapped Hamiltonians that represent invertible topological phases. If we fix a point of $\Ha_{\B}$, then we could \lq\lq identify" this space with the orbit of the stacking action on this point.  This action of the space $E_{d=2}$ on $\Ha_{\B}$ by stacking leads to another fibration as follows.  In general for each point in $\Ha_{\B}$ and the action of  $E_{d=2}$ gives rise to an orbit.  The quotient $\M_\B$ of $\Ha_{B}$ by the $E_{d=2}$ action is defined by declaring each such orbit as a single point in the quotient. It is not yet clear that such mathematical spaces can be defined rigorously to form a fibration: $E_{d=2}\rightarrow \Ha_{\B} \rightarrow \M_{\B}$. Since every space in our discussion is only defined up to homotopy, one potential definition of this fibration is to use a general construction in homotopy theory: every continuous map $f: A\rightarrow B$ can be turned into a fibration.  In this paper, we are mainly interested in the homotopy type and cohomology groups of $\M_\B$.  

It has been conjectured in \cite{Kitaev2011} that short-ranged entangled phases form a $\Omega$-spectrum $\{F_d\}$.  Hence all homotopy groups of $F_d$ for any $d$ can be deduced from those of $F_0$ and $\pi_0(F_d)$.  The space $F_0$ is $\C P^\infty$, and for $F_2$, $\pi_i(F_2)$ is $\mathbb{Z}$ for $i=0,4$ and all other higher homotopy groups vanish.  Hence each connected component of $F_2$ is a Eilenberg-MacLane space $K(\mathbb{Z},4)$.

There is another subtlety about the classifying space $B\uuPic\B$ of categorical groups.  The third homotopy group of $B\uuPic\B$ is isomorphic to the group $\mathbb{C}^\times$ of non-zero complex numbers independent of $\B$ \cite{EtingofNOM2009}. It follows that for $\B$ is trivial, then $B\uuPic\B$ is an Eilenberg-Maclane space $K(\mathbb{C}^\times,3)$.  On the other hand, the connected components of the space of short-ranged entangled phases in two spacial dimensions should be $K(\mathbb{Z},4)$.  It is not clear how these two results are related. But we know even for the ordinary classifying space $BG$ for infinite groups $G$, $BG$ could be different for the different choices of topology of $G$, {\it e.g.} discrete or continuous topologies \cite{stasheff78}.  It is not inconceivable that for some choice of topology of $\mathbb{C}^\times$ and some construction of classifying spaces of categorical groups that $K(\mathbb{C}^\times,3)$ is homotopic to $K(\mathbb{Z},4)$ similar to $K(U(1),1)\simeq BU(1)$, $BU(1)\simeq K(\mathbb{Z},2)$, or some other spaces.\footnote{The equalities hold under suitable choices of topology, using $BU(1)=\mathbb{C}\mathbb{P}^\infty=K(\mathbb{Z},2)$ and $BG=K(G,1)$ for $G=U(1)$ with discrete topology.}

\section{Topology of moduli spaces}
\label{sec:topmoduli}
\subsection{Homotopy types of moduli spaces}

The homotopy type of the moduli space $\M_\B$ for an anyon model $\B$ is conjectured to be the same as $B\uuPic\B$--the classifying space of the categorical symmetry $2$-group of $\B$.  In the following sections, we will provide evidence to this conjecture.    

\subsubsection{Symmetries of topological order and classifying spaces}

Ordinary symmetries are well modelled by groups.  But for anyon models, high categorical groups are better to capture the full extent of symmetries so physical phenomena such as symmetry fractionalization, symmetry defects, and gauging can be appropriately formulated (see {\it e.g.} \cite{symmetry19}).  The high categorical symmetry group of an anyon model $\B$ is the the categorical 2-group $\uuAut\B$, which is isomorphic to $\uuPic\B$.  Analogous to groups, categorical groups have also classifying spaces and we are interested in the classifying space $B\uuPic\B$.  The homotopy groups of $B\uuPic\B$ is calculated in \cite{EtingofNOM2009}: the fundamental group $\Pi_1(\B)$ of $B\uuPic\B$ is isomorphic to the group $\Aut\B$ of $\B$. The second homotopy group $\Pi_2(\B)$ is isomorphic to the group of abelain anyons of $\B$, which is the same as $1$-form symmetries of $\B$.  The third homotopy group $\Pi_3(\B)$ is isomorphic to the group of non-zero complex numbers $\C^\times$.  All other higher homotopy groups vanish.

\subsubsection{Postnikov tower of $B\uuPic\B$}

One way to understand the homotopy type of a space is to use the Postnikov tower.  If the conjecture that $\M_\B$ is the classifying space of $\uuPic\B$ holds, then it follows from the homotopy groups of $B\uuPic\B$ that $\M_\B$ is the twisted product of several Eilenberg-Maclane spaces: $B\uuPic\B \simeq K(\Pi_1(\B), 1)\tilde{\times} K(\Pi_2(\B),2) \tilde{\times} K(\Pi_3(\B),3)$.

\subsection{Cohomology groups of moduli spaces}
\label{sec:cohomology}

Characteristic classes are important cohomology classes that measure the twisting of bundles.  Chern-Weil theory revealed the deep connection between those classes and gauge fields and curvatures.  The hypothetical fibration $E_{d=2}\rightarrow \Ha_{\B} \rightarrow \M_{\B}$ suggests a formal analogy:  if we treat this fibration as a universal bundle, then for any continuous map $f: X\rightarrow \M$ for some space $X$, we would be interested in the "characteristic classes" $\{f^*\omega \in H^*(X)\}, \omega \in H^*(\M)$.

One interesting result of our paper is the construction of such cohomology classes. We will use effective field theories as well as exactly solvable lattice Hamiltonian models.  Our cohomology classes lead to non-trivial information of the $k$-invariants of the Postnokov tower $B\uuPic\B \simeq K(\Pi_1(\B), 1)\tilde{\times} K(\Pi_2(\B),2) \tilde{\times} K(\Pi_3(\B),3)$.

For instance, in Section \ref{sec:EFT} and Section \ref{sec:examples}, we will give examples of family of gapped systems in 2+1d where the $H^4$ cohomology of the moduli space is given by the cup product of the generators in the $H^2$ cohomology, and thus admits a ring structure.
In Section \ref{sec:fermionicfamily} we will give an example with non-trivial Postnikov class.

\section{Effective field theory approach}
\label{sec:EFT}

In the sequel, we will study families of gapped system using the low energy effective field theory. We will assume the gapped system can be described at low energy by a topological quantum field theory (TQFT).  As explained earlier, the universal topological properties in such a field theory will be encoded by some topological order, which is essentially equivalent to the low energy TQFT.  A Hamiltonian realization of the topological order in one sense is the same as a lattice model for the corresponding TQFT. Hence a field theory approach is essentially equivalent to a Hamiltonian approach, though in practice the translation between each other is not always straightforward.

To study the dependence of the effective theory on the parameter in the microscopic gapped system, a useful tool is promoting the parameter to be spacetime-dependent background field:
\begin{equation}
    \lambda: \quad X\rightarrow M~,
\end{equation}
where $M$ is the parameter space, and $X$ is the spacetime manifold. We will characterize the low energy theory by the response under such a probe background field. The partition function of the effective theory depends on these background field.

\subsection{Family of gapped systems through higher-form symmetries}

Let us focus on family of 2+1d bosonic gapped quantum field theories parameterized by $M$, described by action $S[\lambda]$ of parameter $\lambda\in M$.
At low energy, they flow to the same topological quantum field theory  ${\cal B}$ ``enriched" by $M$. 
In the infrared, there can be emergent one-form symmetry ${\cal A}$ generated by the set of Abelian anyons, and ordinary symmetry that permutes the line operators in the TQFT.
We can describe $S[\lambda]$ using the emergent one-form symmetry, by similar method in \cite{Benini:2018reh}.
In \cite{Benini:2018reh}, the low energy theory is enriched with a global 0-form symmetry $G$, by
\begin{itemize}
    \item Permuting the operators in the low energy theory, such as permuting the line operators corresponding to the particle excitations. This can be described by a map $\gamma:G\rightarrow \text{Aut}({\cal B})$.

    \item Decorating various junctions of the generators of 0-form symmetry with the generator of the emergent one-form symmetry (see also {\it e.g.} \cite{symmetry19}). Such decoration can be described by activating a background for the emergent one-form symmetry in terms of the background of the 0-form symmetry \cite{Benini:2018reh}.
    In the case that the one-form symmetry only emerges at low energy, with symmetry group ${\cal A}$, this can be described by a two-cocycle $\eta\in H^2(BG,{\cal A})$: at the junction of 0-form symmetry defects $g_1,g_2,g_1g_2\in G$, there sits the generator of one-form symmetry that for the element $\eta(g_1,g_2)$. As a consequence, the line operators that transform under the one-form symmetry element $\eta(g_1,g_2)$ carry projective representation of $G$, since there is additional phase arise from encircling such line operators around the junction of $g_1,g_2,g_1g_2\in G$, given by braiding the line operators with the generator of the one-form symmetry $\eta(g_1,g_2)$ sitting at the junction.    
    Denote the background for the one-form symmetry by $B_2$, and the background for the 0-form symmetry by $A$, then the cocycle corresponds to activating the background $B_2=A^*\eta$, {\it i.e.} the pullback of $\eta$ by $A:X\rightarrow BG$.

    \item Stacking an invertible phase with $G$ symmetry on the system. For bosonic theories, we can describe such phases by $H^4(BG,\mathbb{Z})$.

\end{itemize}
In the following, we will describe how the low energy theory depends on $\lambda$ in a similar way, where $BG$ is replaced by the target space $M$.

Let us denote the backgrounds for the 0-form symmetry $\text{Aut}({\cal B})$ by $B_1$, and background for the one-form symmetry ${\cal A}$ by $B_2$, and vary the parameter on $M$ over spacetime $X$ to be spacetime-dependent background fields $\lambda:X\rightarrow M$.

We will consider the following class of TQFT enriched by $M$ via the symmetries in the TQFT, by turning on the background fields
\begin{equation}
    B_1=\gamma_M\circ \lambda,\quad B_2=\lambda^*\eta_M~,
\end{equation}
where the theory enriched by $M$ can be described by
\begin{itemize}
    \item Classifying map $\gamma_M : M\rightarrow  B\text{Aut}({\cal B})$.
    
    \item $\eta_M\in C^2(M,{\cal A})$.
    
    \item Torsional Berry phase $H^3(M,U(1))$.
    \item Non-torsion Berry phase $H^4(M,\mathbb{Z})$.
    
\end{itemize}
The relation between the background $B_1,B_2$ and $\lambda$ means that across coordinate patches on $M$, one needs to perform suitable 0-form and one-form gauge transformations.

When the theory has multiple vacua on two-sphere, labelled by the value of topological point operators that generate two-form symmetry ${\cal A}^{(2)}$, we also have $\nu_M\in C^3(M,{\cal A}^{(2)})$, which activates a background $B_3=\lambda^*\nu_M$ for the two-form symmetry.
In the following, we will focus on the case that the theory has unique vacuum on two-sphere, and thus ${\cal A}^{(2)}$ is trivial.

In general, the one-form and 0-form symmetries in the TQFT can combine into a two-group, 
\begin{equation}
    d_{B_1}B_2= \beta(B_1),\quad \beta\in H^3(B\text{Aut}({\cal B}),{\cal A})~,
\end{equation}
where the twist differential uses the action $\text{Aut}({\cal B})$ on ${\cal A}$.
Examples are quantum double theory of Dihedral or Quaternion groups of order 16, they have $\mathbb{Z}_2$ center one-form symmetry and $\mathbb{Z}_2$ outer automorphism 0-form symmetry, which mixes into a two-group, with Postnikov class given by the obstruction to group extension of $\mathbb{Z}_2$ by the Dihedral or Quaternion groups, which is the non-trivial element in $H^3(B\mathbb{Z}_2,\mathbb{Z}_2)=\mathbb{Z}_2$. Then the theory enriched by ${\cal M}$ can have obstruction
\begin{equation}
 \gamma_M ^*\beta\in H^3(M,{\cal A}),\quad d_{\gamma_M }\eta_M =\gamma_M ^*\beta~.
\end{equation}
We will give an example in Section \ref{sec:fermionicfamily}.
In general, the symmetries can have an 't Hooft anomaly. Denote the two-group by $\mathbb{G}$, the anomaly is $\omega\in H^5(B\mathbb{G},\mathbb{Z})$, which for finite groups reduces to the $H^4$ obstruction $H^4(B\mathbb{G},U(1))$.
Then the theory enriched by $M$ has the obstruction 
\begin{equation}
    \lambda^*\omega\in H^5(M,\mathbb{Z})~.
\end{equation}
This obstruction implies that the partition function of the theory cannot be defined as a function on the parameter space $M$ independent of the coordinate on $M$.
It is the analogue of the sigma model anomaly discussed in \cite{Moore:1984ws}, where the presence of fermions implies that the effective action that depends only on the sigma model field cannot be defined as a function on the target space in an coordinate-independent way. In \cite{Moore:1984ws}, the sigma model field is dynamical, and such anomaly is an inconsistency of the theory; here, the sigma model field $\lambda:X\rightarrow M$ is a background field, and the theory is well-defined. Such an anomaly can be realized by topological order instead of by fermions.

We remark that in some cases, the underlying topological order is related to an invertible phase by condensation of anyons. 
If such condensation can be described in the Hamiltonian, then we can relate the family of gapped systems with topological order to the family of gapped systems describing invertible topological phases; the latter are characterized by the Berry phase as discussed in \cite{Kapustin:2020eby,Kapustin:2020mkl,Hsin:2020cgg}. For instance, in \cite{Aasen:2022cdu} the family of gapped Hamiltonians with $\mathbb{Z}_2$ gauge theory topological order, protected by the fundamental group of the parameter space $S^1$, can be obtained from family of free fermion Hamiltonian models by gauging the fermion parity symmetry in the latter Hamiltonian models, or conversely from the fermion condensation in the models with $\mathbb{Z}_2$ topological order.
However, not every topological order can be related to invertible phases by condensation of topological defects, such as $SU(2)_6$, which has chiral central charge $c_-=9/4$ that cannot match the chiral central charge of any invertible phases in 2+1d. Thus the moduli space of the family of gapped Hamiltonians with the same topological order includes and is richer than the family of invertible phases.

\subsection{Example: family of gapped systems realizing $\pi_2=\text{Abelian anyons}$}

Let us describe a construction of family of 2+1d gapped systems parameterized by $M$, that flow to the same TQFT ``enriched by $M$",
with discrete choices classified by $H^2(M,{\cal A})$, where any finite Abelian group can be written as ${\cal A}=\prod \mathbb{Z}_{N_i}$. 
The construction uses higher Thouless pump:
\begin{itemize}
    \item[(1)] 

We start with a family of gapped system with a unique vacuum in $(d+1)$ dimension spacetime, with $\mathbb{Z}_N$ symmetry (for instance, it can be the subgroup of a $U(1)$ symmetry) parameterized by $M$, there can be higher Thouless pump for $\mathbb{Z}_N$ charge described by $\omega_{d}\in H^{d}( M,\mathbb{Z}_N)$ and the effective action: \cite{Hsin:2020cgg,Kapustin:2020mkl}
\begin{equation}
    \int A\cup \omega_d~,
\end{equation}
where $A$ is the background gauge field for the $\mathbb{Z}_N$ symmetry.
The effective action means that if we vary the parameter ${\cal M}$ in spacetime such that $\oint \omega_d\neq 0$ on a codimension one closed domain wall, then there is a $\mathbb{Z}_N$ charge inside the wall. The Thouless pump can be generalized to $n$-form symmetry, where $A$ is replaced by $(n+1)$-form gauge field, and $\omega_d$ is replaced by $\omega_{d-n}$ of degree $(d-n)$.

For gapped lattice Hamiltonian models with $U(1)$ onsite 0-form symmetry, $\omega_d$ is computed in \cite{Kapustin:2020mkl}.

\item[(2)] 

We then gauge $\mathbb{Z}_N$ symmetry to obtain family of (non-invertible) gapped systems with $\mathbb{Z}_N$ gauge theory topological order\footnote{
More generally, we can couple the system to particles in $\mathbb{Z}_N$ gauge theory that can have non-trivial statistics.
}, parametrized by
\begin{equation}
H^{d}(M,\mathbb{Z}_N)~.   
\end{equation}
In lattice model, one can gauge onsite $\mathbb{Z}_N$ symmetry following the method of {\it e.g.} \cite{PhysRevB.86.115109,Bhardwaj:2016clt,Shirley:2018vtc,Tsui:2019ykk,PhysRevB.87.125114}.

\end{itemize}

\begin{figure}[t]
  \centering
    \includegraphics[width=0.3\textwidth]{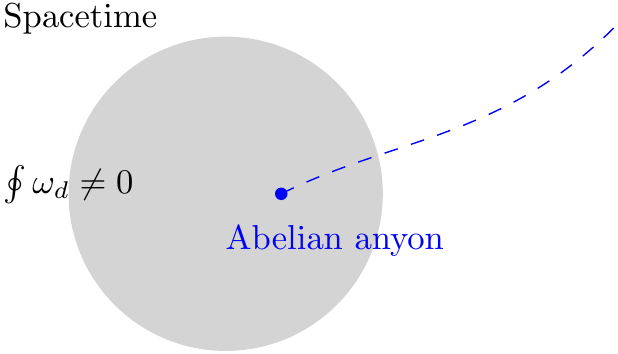}  
    \caption{Varying the parameter over spacetime can trap particles. 
    }\label{fig:anyon}
\end{figure}

As an application, in such family of systems in 2+1d one can trap anyons by varying the parameters over spacetime such that $\int \omega_{d}\neq 0$, see Figure \ref{fig:anyon}. This can have experimental applications. Instead, we can also couple the symmetry charges to anomalous one-form symmetry.

We will give example in the next section with $M=S^2,T^2$ and $M=S^2\times S^2$. They can exhibit the following features:
\begin{itemize}
    \item Fractional higher Berry phase. While the higher Berry phase in invertible gapped system is quantized \cite{Hsin:2020cgg,Kapustin:2020eby}, the family of non-invertible gapped systems can produce fractional higher Berry phase. If we probe the system by varying the parameters in space, it is a fractional response. This can have experimental applications.
Examples of such fractional response is discussed in Section \ref{sec:examplefracBerryS2xS2} and Section \ref{sec:exampleT4fracBerryphase}.

    \item For a family of gapped systems that flow to ${\cal B}=\mathbb{Z}_N$ untwisted gauge theory in 2+1d, the higher Berry phase is given by cup product of the elements in $H^2(M,{\cal A})$, where ${\cal A}=\mathbb{Z}_N^e\times\mathbb{Z}_N^m$, and
    the coefficient of the cup product is the braiding of anyons in $\mathbb{Z}_N$ toric code
    \begin{equation}
        H^2(M,{\cal A})\times H^2(M,{\cal A})\rightarrow H^4(M,\mathbb{R}/\mathbb{Z}),\quad \eta\in H^2(M,{\cal A})\mapsto \theta(\eta)~,
    \end{equation}
    where $\theta(\eta)$ is the topological spin of the Abelian anyon $\eta$.  $\theta(\eta)$ is an $N$-torsion element, and can be written as $1/N$  fractional higher Berry phase.
Examples of such structures are discussed in Section \ref{sec:examplecupproduct}.

    \item Some higher Berry phase can be absorbed by the gapped system. In other words, some invertible systems with non-trivial higher Berry phase can be smoothly connected to system with trivial Berry phase when it is stacked with family of non-invertible gapped systems. 
    One way to understand this is that when varying the parameter over space such that there is a single particle sitting on the space with worldline along the time direction, the Hilbert space is empty.\footnote{
    Similar phenomena in symmetry fractionalization is observed in \cite{Delmastro:2022pfo}. Examples of symmetry protected topological phases become trivial when stacked with symmetry enriched topological phases are discussed in 
    \cite{PhysRevX.6.011034,Hsin:2019fhf,Choi:2022zal}.
    }
Examples of such ``trivialized" Berry phase are discussed in Section \ref{sec:exampletrivializedBerry}.
    
\end{itemize}

\paragraph{Application: manipulation of Abelian anyons}

Let us consider the family with background $B_2=\lambda^*\eta_M$ for one-form symmetry ${\cal A}$. Denote the 2d space coordinate by $x,y$ and time coordinate by $t$. Let us vary the parameter over space $\lambda=\lambda(x,y)$, then $B_2(x,y)=\lambda(x,y)^*\eta_M $
is a two-form in space with value in ${\cal A}$. We would like to choose a configuration of $\lambda=\lambda(x,y)$ such that the above two-form $B_2(x,y)=\lambda(x,y)^*\eta_M$ has a localized profile:
\begin{equation}
    B_2(x,y)=\lambda(x,y)^*\eta_M =\sum_i \alpha_i \delta(x-x_i)\delta(y-y_i)dxdy~,
\end{equation}
for some Abelian anyons $\alpha_i\subset  {\cal A}$ localized at the space locations $(x,y)=(x_i,y_i)$.

For instance, if the parameter space is $M=T^2$, with coordinate $(\lambda_1,\lambda_2)\sim (\lambda_1+1,\lambda_2)\sim (\lambda_1,\lambda_2+1)\in \mathbb{R}^2/\mathbb{Z}^2= T^2$, then $H^2(M,{\cal A})\cong {\cal A}$, with $\lambda^*\eta_M=\alpha d\lambda_1 d\lambda_2$ for $\alpha\in {\cal A}$, and we can take
\begin{equation}
    \lambda(x,y)=(h(x-x_1)-h(x-x_2),h(y-y_1)-h(y-y_2))~,
\end{equation}
where $h$ is the step function. This gives 
\begin{equation}
    B_2= \sum_{i=1}^2 \alpha_i \delta(x-x_i)\delta (y-y_i) dxdy,\quad \alpha_1=-\alpha_2=\alpha\in{\cal A}~.
\end{equation}
In other words, by abruptly changing the parameter from (0,0) to (1,1) at some points $\{(x_i,y_i)\}_{i=1,2}$, we effectively realize a potential that traps a pair of Abelian anyons $\alpha,-\alpha$ at these points. We can also consider a smeared version of the step function over some finite range of length $\ell_\text{localize}$, then the Abelian anyons will be localized within the range $\ell_\text{localize}$.

We can also make $x_i=x_i(t),y_i=y_i(t)$ to have time dependence to transport the Abelian anyons along some trajectory in space. For instance, we can create a pair of Abelian anyons at some point, transporting them along some non-contractible cycle in space (such as in a direction that has periodic boundary condition), then annihilate the pair of anyons. This realizes a non-trivial logical gate on the ground state subspace.

\subsection{Example: family of gapped systems with $H^3$ obstruction}
\label{sec:fermionicfamily}

Let us describe a family of gapped systems in 2+1d with parameter space $M$, such that the family of systems can have an $H^3$ obstruction.
The low energy theories have $\mathbb{Z}_2$ topological order with fermion particles that belong to the topological order studied in \cite{Kitaev:2005hzj}. The system has parameter space $M$.
We will ``enrich" the topological order with the parameter space by the emergent higher-form symmetry, similar to the enriching the $\mathbb{Z}_2$ topological order with ordinary symmetry $G$ as discussed in 
\cite{PhysRevB.105.235143} (see also \cite{PhysRevB.90.115141}).

Let us focus on the gapped systems with integer chiral central charge $c_-$, which includes the case $c_-=0$.

The $\mathbb{Z}_2$ topological order has emergent $\mathbb{Z}_2$ electromagnetic duality 0-form symmetry, and one-form symmetries generated by the electric and magnetic particles. For even $c_-$, the particle spectrum forms a $\mathbb{Z}_2\times\mathbb{Z}_2$ fusion algebra, while for odd $c_-$ the particle spectrum is a $\mathbb{Z}_4$ fusion algebra. We will omit the time-reversal symmetry in the discussion.
Denote the background of the 0-form symmetry by $B_1$, and the background of
and one-form symmetries by $B_2,B_2'$, they obey  the following relations:
\cite{Hsin:2020nts,PhysRevB.105.235143}
\begin{equation}\label{eqn:Z2extension}
    dB_1=0,\quad dB_2=B_1\cup B_2'+ c_- B_2'\cup_1 B_2'~.
\end{equation}
Such equation describes the extension of $\mathbb{Z}_2$ one-form symmetry by $\mathbb{Z}_2$ one-form symmetry depending on $c_-$ mod 2, with automorphism action given by the electromagnetic duality 0-form symmetry. The 0-form and one-form symmetries have an anomaly, described by the bulk symmetry protected topological phase with the 0-form and one-form symmetries, and the effective action:\cite{PhysRevB.105.235143} 
\begin{align}
    S_\text{bulk}&=\pi\int\left( {\cal P}(B_2)+B_2\cup B_2'+
    c_-(B_1\cup B_2')\cup_2(B_2'\cup_1 B_2')+
    \zeta(B_1,B_2')\right)\cr
    &\qquad +\frac{\pi}{2}\int \tilde B_1\cup \tilde B_1\cup \tilde B_2
    +\frac{\pi c_-}{4}\int {\cal P}(B_2')
    ~,
\end{align}
where tilde denotes the lift of the $\mathbb{Z}_2$ background fields to $\mathbb{Z}_4$ with value $0,1$, ${\cal P}(x)=x\cup x-x\cup_1 dx$ is the Pontryagin square,
and $\zeta$ is defined as
$\zeta(B_1,B_2')=B_1\cup\left((B_1\cup B_2')\cup_2 B_2'+B_1\cup B_2'\right)$.

\subsubsection{Family without $H^3$ obstruction}

We can obtain a family of gapped system with such topological order protected by $\pi_1(M),\pi_2(M)$ given by
  promoting $\lambda$ to be spacetime-dependent background fields, and substituting $B_1=\lambda^*n_1,B_2=\lambda^*n_2,B_2'=\lambda^*\omega_2$ in equation (\ref{eqn:Z2extension}), with
\begin{align}
    &n_2\in C^2(M,\mathbb{Z}_2),\quad \omega\in H^2(M,\mathbb{Z}_2),\quad n_1\in H^1(M,\mathbb{Z}_2)\cr
    &dn_1=0,\quad dn_2=n_1\cup \omega_2+c_-\omega_2\cup_1\omega_2~.
\end{align}
We remark that for the special case $M=BG_b$ with group $G_b$, these data describes invertible fermionic phases with symmetry given by extension of $G_b$ by the $\mathbb{Z}_2^f$ fermion parity symmetry; equivalently, they describe the ``bosonic shadow" $\mathbb{Z}_2$ gauge theory with fermion Wilson lines and enriched with $G_b$ 0-form symmetry, related to the invertible fermionic phases by summing over the spin structures \cite{PhysRevB.105.235143}.

If such solutions for $n_1,n_2,\omega_2$ exist, then the family does not have an $H^3$ obstruction. For instance, if $M=T^2$, with parameter $\lambda=(\lambda_1,\lambda_2)$ for real-valued $\lambda_i\sim \lambda_i+1$, we can take $n_1=d\lambda_1$ mod 2, $n_2=\omega_2=d\lambda_1 d\lambda_2$ mod 2. Then $dn_2=0$, which is consistent with $n_1\omega_2=0$, and $\omega_2\cup_1 \omega_2=d\tilde \omega_2/2=0$ where $\tilde \omega_2=d\lambda_1d\lambda_2$ mod 4 is a lift of $\mathbb{Z}_2$-valued $\omega_2$ to $\mathbb{Z}_4$ value.

The family also has Berry phase, given by the anomaly of the 0-form and one-form symmetries:
\begin{align}
  &  S_\text{bulk}
  =2\pi \int \lambda^*\Omega_4~,\cr
&\Omega_4=    \frac{1}{2}\left( {\cal P}(n_2)+n_2\cup \omega_2+
    c_-(n_1\cup \omega_2)\cup_2(\omega_2\cup_1 \omega_2)+
    \zeta(n_1,\omega_2)\right)\cr 
    &\qquad +\frac{1}{4} \tilde n_1\cup \tilde n_1\cup \tilde n_2
    +\frac{c_-}{8}{\cal P}(\omega_2)
    ~.
\end{align}

\subsubsection{Family with $H^3$ obstruction}

If the data $(n_1,n_2,\omega_2)$ does not obey the equation $dn_2=n_1\cup \omega_2+c_-\omega_2\cup_1\omega_2$, the family has an $H^3$ obstruction.\footnote{
For given $n_1,\omega_2$ there can be no solution for $n_2$. For instance, take $M=S^1\times S^2$, and $n_1$ being the integral volume form on $S^1$ modulo 2, and $\omega_2$ being the integral volume form on $S^2$ modulo 2, then $n_1\cup \omega_2+c_-\omega_2\cup_1 \omega_2$ is the integral volume form on $S^1\times S^2$ modulo 2, which is closed but not exact in the cohomology with $\mathbb{Z}_2$ coefficient, and thus the solution for $n_2$ does not exist.
}
For instance, if we set $B_2'=\lambda^*\omega_2$, $B_1=\lambda^*n_1$, and $n_2=0$, then we need to turn on non-trivial $B_2$:
\begin{equation}
    dB_2 = \lambda^*\Omega_3,\quad \Omega_3=n_1\cup \omega_2+c_-\omega_2\cup_1\omega_2~.
\end{equation}
The equation implies that the point that intersects the region  with non-trivial $\int \lambda^*\Omega_3$ emits an Abelian anyon that generates the one-form symmetry whose background is $B_2$.

We remark that this is an analogue of the property that different (relative) fractionalization classes can change the higher-group symmetry, see {\it e.g.} Appendix E in \cite{Delmastro:2022pfo}.

\section{Example: gapped systems with $\mathbb{Z}_N$ topological order}
\label{sec:examples}

\subsection{Parameter space $S^2$}
\label{sec:exampleS2}

We will focus on 2+1d spacetime.
Let us consider $U(1)$ gauge theory with two complex scalars $\phi_i$, $i=1,2$, of charge one, coupled to $\mathbb{Z}_N$ gauge theory described by $U(1)\times U(1)$ BF theory:
\begin{equation}
   \frac{1}{2e^2}(da)^2+ |D_a\phi_i|^2+V(\phi)+\frac{da}{2\pi}(\alpha u_e+\beta u_m)+\frac{N}{2\pi}u_edu_m~,
\end{equation}
where $a,u_e,u_m$ are $U(1)$ gauge fields, and $u_e,u_m$ describe $\mathbb{Z}_N$ gauge theory, with the particles with electric and magnetic charges $q_e,q_m$ described by $e^{i\int (q_e u_e+q_mu_m)}$. $\alpha,\beta$ are integers.
The potential is
\begin{equation}
    V(\phi)=- \sum_{i,j=1,2; a=1,2,3}\phi^* _i \sigma_{ij}^a m^a \phi_j+\sum_i\lambda (\phi_i^*\phi_i)^2~.
\end{equation}
We normalize the mass parameters $m^a$ as
\begin{equation}
    m_1^2+m_2^2+m_3^2=\mu^4~.
\end{equation}
For nonzero $\mu$, the scalars condense, with the expectation value depending on the mass parameters $m^a$, and thus Higgsing the $U(1)$ gauge field $a$,\cite{Hsin:2020cgg}
\begin{equation}
\text{Low energy Higgs phase:}\quad     {da\over2\pi}=\frac{1}{2\pi}\epsilon^{abc}m^adm^b dm^c~,
\end{equation}
which identifies the first Chern class of the $U(1) $ gauge field $a$ on the left with the generator of $H^2(S^2,\mathbb{Z})$ on the right.

The theory has effective action
\begin{equation}
    \frac{1}{2\pi}\epsilon^{abc}m^adm^b dm^c (\alpha u_e+\beta u_m)+\frac{N}{2\pi}u_edu_m~.
\end{equation}

\paragraph{Particle trap}

This implies that if we vary the parameters $m^a$ such that there is nonzero Skyrmion number $\int_{\Sigma} \epsilon_{abc}m^a dm^b dm^c\neq 0$ on space $\Sigma$, there is a particle with charge $(q_e=\alpha,q_m=\beta)$ anyon $e^{i\int(\alpha u_e+\beta u_m)}$. (For more general configuration $m^a$, the worldline of the particle is at the Poincar\'e dual of $\epsilon_{abc}m^adm^bdm^c$.)
Such anyon has spin $\frac{\alpha \beta}{N}$ mod 1.

\paragraph{Family of gapped systems with $\pi_2=\mathbb{Z}_N\times\mathbb{Z}_N$}

The couplings $\alpha,\beta$ describe
\begin{equation}
    (\alpha,\beta)\in H^2(S^2,\mathbb{Z}_N^e\times \mathbb{Z}_N^m)=\mathbb{Z}_N^e\times \mathbb{Z}_N^m~.
\end{equation}

\subsubsection{Candidate lattice Hamiltonian realization}

\begin{figure}[t]
  \centering
    \includegraphics[width=0.6\textwidth]{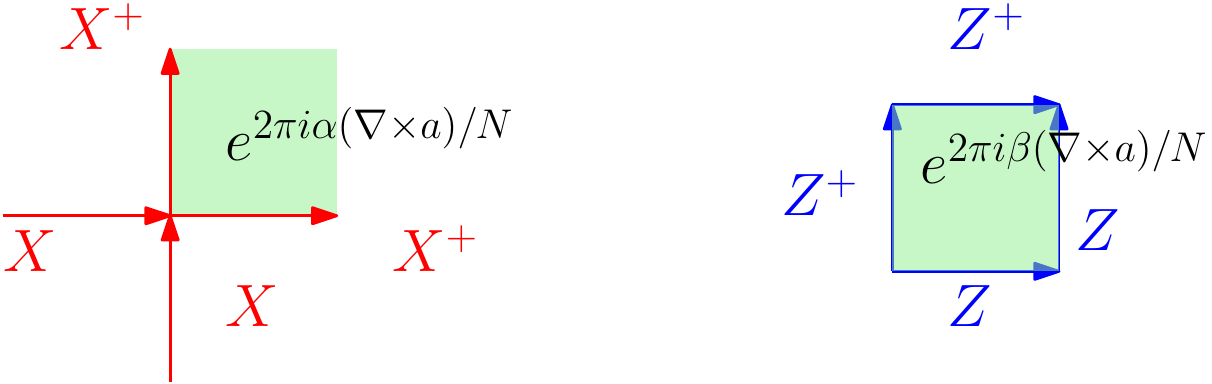}
    \caption{The last two terms in the Hamiltonian (\ref{eqn:latticeAbelianHiggs}), where the left figure is the Gauss law term at each site where the four red edges meet, and the right figure is the flux term at each face surrounded by the four blue edges.
    }\label{fig:LatticeAbelianHiggs}
\end{figure}

Consider square lattice with 2-component complex scalar $z^i_{(x,y)}$, $i=1,2$ at each vertex $(x,y)$,\footnote{
We will also denote the vertex by $\mathbf{x}=(x,y)$, and the unit vector along $x,y$ direction by $\hat x,\hat y$, respectively.
} and gauge field $a_e\in \mathbb{R}/\mathbb{Z}$ and electric field $E_e\in\mathbb{Z}$ at each edge $e$, with the commutation relation $[a_e, E_{e'}/g^2]=i\delta_{e,e'}$.
We also introduce $\mathbb{Z}_N$ degrees of freedom on each edge, acted by the analogue of Pauli matrices ${\cal X},{\cal Z}$ that satisfies ${\cal XZ}={\cal ZX}e^{2\pi i/N}$.
We propose the following Hamiltonian
\begin{align}\label{eqn:latticeAbelianHiggs}
    H=& -\sum_{s=1,2,3}J_s\sum_{\mathbf{x};\,\mu=\hat x,\hat y} \left(z^i_{\mathbf{x}}\sigma_{ij}^s e^{ia_{(\mathbf{x},\mu)}}z^j_{\mathbf{x}+\mu}\right)
    -K\sum_\text{face} e^{i \nabla\times a}
    -g^2 \sum_e e^{iE_e/g^2}{\cal X}_e^{-\beta}{\cal Z}_{L(e)}^{-\alpha }\cr 
    &-\sum_\text{sites} {\cal X}{\cal X}{\cal X}^\dag {\cal X}^\dag e^{\frac{2\pi i\alpha}{N} (\nabla \times a)}
    -\sum_\text{faces} {\cal Z}{\cal Z}{\cal Z}^\dag {\cal Z}^\dag e^{\frac{2\pi i\beta}{N} (\nabla \times a)}+\text{h.c.}~,
\end{align}
where $L(\mathbf{x},\hat x)=(\mathbf{x}-\hat y,\mathbf{x})$, $L(\mathbf{x},\hat y)=(\mathbf{x}-\hat x,\mathbf{x})$, and $\nabla \times a$ is a small Wilson loop of the gauge field $a$ around a square. The last two terms in the Hamiltonian are depicted in Figure \ref{fig:LatticeAbelianHiggs}.
The last two terms commute with each term of the  Hamiltonian, and thus we do not need to specify their coefficients. If dropping the $a$ dependence in the last two terms, they describe $\mathbb{Z}_N$ toric code.
We will focus on the limit of large $g^2$, where the term depending on $E_e$ can be approximated by $\frac{1}{g^2}\sum E_e^2$, then the last two terms in the first line becomes the kinetic energy of $U(1)$ lattice gauge theory (omitting the ${\cal X},{\cal Z}$ operators).
We will also take $K,J_s$ to be large,
then we expect the theory to be in the Higgs phase, where the scalar condenses and the gauge field does not fluctuate.
Similar lattice Hamiltonian model for Abelian Higgs model without $\mathbb{Z}_N$ gauge field is considered in {\it e.g.} \cite{Bonati:2020jlm}.

\subsubsection{Parameter space $S^2\times S^2$ and fractional higher Berry phase}
\label{sec:examplefracBerryS2xS2}

We can introduce another copy of complex scalars with independent mass terms $m'^a$ that parametrize another $S^2$, and another $U(1)$ gauge field that is Higgsed by the scalar. Let us denote the parameters in the second copy with a prime. Both copies couple to the same $\mathbb{Z}_N$ gauge theory.

The total theory has following effective action for the $\mathbb{Z}_N$ gauge theory:
\begin{equation}
   \frac{1}{2\pi}\epsilon^{abc}\left((\alpha m^adm^b dm^c+\alpha' m'^adm'^b dm'^c)u_e+(\beta m^adm^b dm^c+\beta' m'^adm'^b dm'^c)u_m\right)
+        \frac{N}{2\pi}u_edu_m~.
\end{equation}
This implies that at the place with nonzero $\int (\alpha m^adm^b dm^c+\alpha' m'^adm'^b dm'^c)$ there is an $e$ particle, and
at the place with nonzero $\int (\beta m^adm^b dm^c+\beta' m'^adm'^b dm'^c)$ there is an $m$ particle.
The family is labelled by
\begin{equation}
    (\alpha,\beta;\alpha',\beta')\in H^2(S^2\times S^2,\mathbb{Z}_N^e\times\mathbb{Z}_N^m)=\mathbb{Z}_N^4~.
\end{equation}

\subsubsection{Higher Berry phase given by cup product, and fractoinal Higher Berry phase response}
\label{sec:examplecupproduct}

Let us consider the $\mathbb{Z}_N$ gauge theory with backgrounds $B_e,B_m$ for the $\mathbb{Z}_N\times\mathbb{Z}_N$ electric and magnetic one-form symmetry that acts on the electric and magnetic particles:
\begin{equation}
    \frac{N}{2\pi}u_edu_m+\frac{N}{2\pi} (u_mB_e+u_eB_m)
\end{equation}
The theory is not gauge invariant under the transformation $B_e\rightarrow B_e+d\lambda_e,B_m\rightarrow B_m+d\lambda_m$, $u_e\rightarrow u_e-\lambda_e$, $u_m\rightarrow u_m-\lambda_m$.
We can cure it by introducing a 3+1d bulk,
\begin{equation}
    \frac{N}{2\pi}\int_{4d}(du_e +B_e)(du_m+B_m)=\frac{N}{2\pi}\int_{3d}\left(u_edu_m+ (u_mB_e+u_eB_m)\right)+\frac{N}{2\pi}\int_{4d}B_eB_m~,
\end{equation}
where we used $dB_e=dB_m=0$. The left hand side is gauge invariant, and thus the 3d-4d system on the right hand side is also gauge invariant.

The family of gapped systems given by
\begin{equation}
  B_e=\frac{1}{N}(\beta\epsilon_{abc}m_am_bm_c+\beta'\epsilon_{fgn}m_f'm_g'm_h'),\quad
  B_m= \frac{1}{N}(\alpha\epsilon_{abc}m_am_bm_c+\alpha'\epsilon_{fgn}m_f'm_g'm_h')~.
\end{equation}

The Higher Berry phase is given by cup product
\begin{equation}
    \frac{N}{2\pi}\int_{4d} B_e B_m
    =\frac{1/N}{2\pi}(\alpha\beta'+\alpha'\beta)\epsilon_{abc}\epsilon_{fgh}\int_{4d}   m_adm_b dm_c  m'_fdm'_g dm'_h~.
\end{equation}
In general, this is $1/N$ of the quantized Berry phase in invertible systems.

\subsubsection{Example with trivialized higher Berry phase}
\label{sec:exampletrivializedBerry}

Let us trivially enlarge the parameter space to be $S^2\times M'$ where the above system is constant over $M'$. Take $\alpha=1,\beta=0$. 
This turns on $B_m=\frac{1}{N}\epsilon_{abc}m_adm_bdm_c$.

If we turn on background $B_e$ for the electric one-form symmetry, we find a bulk Thouless pump
 for $S^2$ and $\mathbb{Z}_N$ one-form symmetry acting on the electric particle:
\begin{equation}
    S_\text{4d}=\frac{N}{2\pi}\int B_eB_m=\int_{4d} \frac{1}{2\pi} \epsilon_{abc}m_a dm_b dm_c B_2^e~.
\end{equation}
This implies that if we perform a one-form transformation $B_e\rightarrow B_e+d
\lambda_e$ with
 $\lambda_e\in H^1(M',\mathbb{Z}_N)$, this gives the boundary Berry phase
\begin{equation}
    S_\text{3d}=\int_{3d} \frac{1}{2\pi} \epsilon_{abc}m_a dm_b dm_c \lambda_e~.
\end{equation}
Thus if we stack the system with the above Berry phase, it is the as stacking nothing, since it can be removed by a one-form transformation. In other words, some higher Berry phase can be continuously connected to the trivial Berry phase in the presence of topological order.

For instance, if the parameter space is $M=S^2\times S^1$ and $\lambda_e=(1/N) d\theta$ with $\theta\sim \theta+2\pi$ being the coordinate on $S^1$, then this Berry phase can be removed when staked with the family of $\mathbb{Z}_N$ gauge theory.

\subsection{General topological order in 2+1d}

For a 2+1d TQFT ${\cal B}$ with $\mathbb{Z}_N$ Abelian anyon of spin $p/(2N)$ mod 1 for integer $p\sim p+2N$, we have duality: \cite{Cordova:2017vab}
\begin{equation}
    {\cal B}\quad \longleftrightarrow\quad {{\cal B}\times (\mathbb{Z}_N)_{Np}\over \mathbb{Z}_N}~,
\end{equation}
where the quotient denotes gauging a diagonal one-form symmetry, or condensing the composite boson of an Abelian anyon from ${\cal B}$ and an Abelian anyon from $(\mathbb{Z}_N)_{Np}$, as discussed in {\it e.g.} \cite{Moore:1989yh,Burnell:2017otf,Hsin:2018vcg}.\footnote{The theory $(\mathbb{Z}_N)_k$ is the twisted $\mathbb{Z}_N$ gauge theory that can be described by $U(1)\times U(1)$ Chern-Simons theory as \cite{Maldacena:2001ss,Banks:2010zn,Kapustin:2014gua}
\begin{equation}
    \frac{N}{2\pi}udv+\frac{k}{4\pi}udu~.
\end{equation}
The field redefinition $v\rightarrow v+u$ gives the equivalence relation $k\sim k+2N$.
} On the right hand side, for even $p$ we have $(\mathbb{Z}_N)_{Np}=(\mathbb{Z}_N)_{0}$, which is the $\mathbb{Z}_N$ Toric code.
For odd $p$ we have
$(\mathbb{Z}_N)_{Np}=(\mathbb{Z}_N)_N=U(1)_N\times U(1)_{-N}$, which is the generalization of the double semion model with the semion self-statistics $i$ replaced by $e^{\pi i/N}$.
Thus the Abelian anyons in any theory can be embedded in the product of $(\mathbb{Z}_{N_i})_0$ or $(\mathbb{Z}_{N_i})_{N_i}$ for a set of integers $\{N_i\}$.\footnote{
The same result can be obtained from stablizer model as in \cite{Ellison:2022web}.
}

The duality enables one to couple the TQFT to the Abelian Higgs model using the $\mathbb{Z}_N$ gauge field,
\begin{align}
&
\frac{1}{2e^2}(da)^2+
\sum_{i=1,2}|D_a\phi_i|^2 +\sum_{i,j=1,2}\sum_{a=1,2,3} \phi^*_i\sigma^a_{ij}\phi_j m_s-\lambda\sum_i (\phi^*_i\phi_i)^2+\frac{q}{2\pi}(pu+v)da\cr
&+\frac{N}{2\pi}udv+\frac{Np}{4\pi}udu+\frac{N}{2\pi}(-v) b+S_{{\cal B}}[b]~,
\end{align}
where $b$ denotes the gauge field for gauging the diagonal $\mathbb{Z}_N$ one-form symmetry, 
$S_{{\cal B}}[b]$ denotes the TQFT ${\cal B}$ coupled to $b$, and $q$ is an integer parametrize the coupling.
The above action describes non-trivial family  gapped systems that flow to TQFT ${\cal B}$ with the same topological order parameterized by $q\in H^2(S^2,\mathbb{Z}_N)=\mathbb{Z}_N$.

\paragraph{Example: $SU(2)_4$}

Let us consider the non-Abelian TQFT $SU(2)_4$ with Abelian anyon: it has 5 particles labelled by $SU(2)$ isospin $j=0,\frac{1}{2},1,\cdots,2$, where $j=2$ is a boson that obeys $\mathbb{Z}_2$ fusion rule. Thus $N=2,p=0$. Let us take $q=1$. The theory has action
\begin{align}
&
\int \left(\frac{1}{2e^2}(da)^2+
\sum_{i=1,2}|D_a\phi_i|^2 +\sum_{i,j=1,2}\sum_{a=1,2,3} \phi^*_i\sigma^a_{ij}\phi_j m_s-\lambda\sum_i (\phi^*_i\phi_i)^2\right)\cr
&\quad +\int\left(\frac{2}{2\pi}udv+\frac{1}{2\pi}vda+(-v) w_2^{SO(3)}\right)+S_{SO(3)_2}[w_2^{SO(3)}]~,
\end{align}
where $S_{SO(3)_2}[w_2^{SO(3)}]$ is the action for the $S_{SO(3)_2}$ Chern-Simons term, where the gauge field has second Stiefel-Whitney class $w_2^{SO(3)}$ for the $SO(3)=SU(2)/\mathbb{Z}_2$ gauge bundle,
and the $\mathbb{Z}_2$ gauge theory described by $u,v$ is equivalent to imposing the condition
\begin{equation}
    \frac{da}{2\pi}=w_2^{SO(3)}\quad \text{mod }2~.
\end{equation}
This relation between the $U(1)$ gauge field $a$ and  the $SU(2)/\mathbb{Z}_2$ gauge field implies that they combine into $U(2)=\left(SU(2)\times U(1)\right)/\mathbb{Z}_2$ gauge field.
Thus the theory is the same as $U(2)_{4,0}$ Chern-Simons theory coupled to two complex scalars $\phi_i$ in the one-dimensional representation $(\mathbf{1},\mathbf{1}_2)$ of $U(2)=\left(SU(2)\times U(1)\right)/\mathbb{Z}_2$, with the quartic potential that depends on the mass parameters $m_s$.

\subsection{Lattice model from reduction of free fermion Chern insulator}
\label{sec:exampleChernT2}

We can construct lattice example with higher Thouless pump by reducing 4+1d Chern Insulator on torus down to 2+1d and treating the components of the extra spatial momenta $k^z=\lambda^z,k^w=\lambda^w$ as parameter $M=T^2$, following \cite{Kapustin:2020mkl}.
The $U(1)$ symmetry is the fermion number symmetry.
The Block wave function describes a vector bundle over $T^2_{k^x,k^y}\times {\cal M}$ with $M=T^2_{\lambda^z,\lambda^w}$, and the two-form $\omega_2$ that describes the higher Thouless pump is given by the integral of the curvature of the connection on this vector bundle over $T^2_{k^x,k^y}$ \cite{Kapustin:2020mkl}.

Then we couple the theory to $\mathbb{Z}_N$ toric code model using the $\mathbb{Z}_N\subset U(1)$ subgroup symmetry to obtain a family of $\mathbb{Z}_N$ gauge theories.
This gives explicit examples of family of exactly solvable Hamiltonian lattice models 
for non-trivial $K(\Pi_2({\cal B}),2)$ factor of $B\text{Pic}({\cal B})$ for $\mathbb{Z}_N$ gauge theory topological order.

\subsubsection{Higher Thouless pump with free fermions and parameter space $T^2$}

The model is defined on square lattice, and it has translation symmetry. The Hamiltonian written in momentum space is:  \cite{Kapustin:2020mkl}
\begin{equation}
    H=\sum_{k_x,k_y}\psi^\dag_\mathbf{k} d_a(\mathbf{k},\lambda^i)\Gamma^a\psi_\mathbf{k}~,
\end{equation}
where $\lambda^i=(\lambda^z,\lambda^w)$ are the momenta in higher dimensions that are reduced, $\Gamma^a$ is the gamma matrices in five spacetime dimension, and $\mathbf{k}=(k^x,k^y)$. The coefficient $d^a$ is
\begin{equation}
    d_a(\mathbf{k},\lambda^i)=\left(m+c(\cos k^x+\cos k^y+\cos\lambda^z+\cos\lambda^w),\sin k^x,\sin k^y,\sin \lambda^z,\sin\lambda^w\right)~,
\end{equation}
where we can take $-4c<m<-2c$, then
 $\omega_2$ is given by the volume form on parameter space $T^2$, with the integral over the entire parameter space equal one \cite{PhysRevB.78.195424}.

We can rewrite the Hamiltonian in real space:
\begin{align}
    H=&\sum \psi^\dag(x,y)\left(\Gamma^1
    (m+c(\cos\lambda^z+\cos\lambda^w))+\Gamma^4 \sin\lambda^z+\Gamma^5\sin\lambda^w
    \right)\psi(x,y)\cr 
    &+\frac{c}{2}\sum \left(\psi(x+1,y)^\dag +\psi^\dag(x-1,y)+
    \psi(x,y+1)^\dag +\psi^\dag(x,y-1)\right)\Gamma^1 \psi(x,y)\cr
    &+\frac{1}{2i}\sum  \left(\psi(x+1,y)^\dag -\psi^\dag(x-1,y)\right)\Gamma^2 \psi(x,y)\cr
    &+\frac{1}{2i}\sum  \left(\psi(x,y+1)^\dag -\psi^\dag(x,y-1)\right)\Gamma^3 \psi(x,y)~.
\end{align}

This is a family of gapped Hamiltonians with unique ground state and $U(1)$ symmetry and higher Thouless pump.
 The local $U(1)$ symmetry transformation is
 \begin{equation}
     U(x,y)=\exp \left(i\theta \psi^\dag(x,y)\psi(x,y)\right),\quad U(x,y)\psi(x,y)U^\dag(x,y)=e^{i\theta}\psi(x,y)~.
 \end{equation}

\subsubsection{Coupling to $\mathbb{Z}_N$ toric code: family of $\mathbb{Z}_N$ topological order}

Let us gauge the $\mathbb{Z}_N\subset U(1)$ subgroup symmetry of the model. 
We introduce $\mathbb{Z}_N$ degrees of freedom on each link, acted on by
the $\mathbb{Z}_N$ analogue of Pauli matrices ${\cal X},{\cal Z}$, with the commutation relation ${\cal XZ}={\cal ZX}e^{2\pi i/N}$.

Without fermion fields, pure $\mathbb{Z}_N$ gauge theory is described by toric code\footnote{
Here it is an untwisted $\mathbb{Z}_N$ quantum double, corresponding to ``minimally'' gauging the $\mathbb{Z}_N$ symmetry without stacking extra symmetry protected topological phases.
}
\begin{equation}
    H_\text{toric code}=H_\text{Gauss law}+H_\text{flux},\quad H_\text{Gauss law}=-\sum_\text{vertices} {\cal X}{\cal X}{\cal X}^\dag{\cal X}^\dag,\quad H_\text{flux}=-\sum_\text{faces} {\cal Z}{\cal Z}{\cal Z}^\dag{\cal Z}^\dag~. 
\end{equation}
Explicitly, the Gauss law term is product of four ${\cal X}$ met at vertex $(x,y)$, while the flux term is product of four ${\cal Z}$ on the boundary edge of the square whose lower left corner is $(x,y)$:
\begin{align}
    &H_\text{Gauss law}= 
-\sum_{(x,y)}    {\cal X}_{(x-1,y),(x,y)}{\cal X}^\dag_{(x,y),(x+1,y)}{\cal X}_{(x,y-1),(x,y)}{\cal X}^\dag_{(x,y),(x,y+1)}
    \cr
     &H_\text{flux}=-\sum_{(x,y)} {\cal Z}_{(x,y),(x+1,y)} {\cal Z}_{(x,y+1),(x+1,y+1)} {\cal Z}_{(x,y+1),(x+1,y+1)}^\dag {\cal Z}_{(x,y),(x,y+1)}^\dag~.
\end{align}

In the presence of fermion fields, we need to modify the above terms and also the Hamiltonian of the fermion fields such that the modified fermion Hamiltonian commutes with both the modified Gauss law terms and the modified flux terms.
The modifications are labeled by
\begin{equation}
    (\alpha,\beta)\in \mathbb{Z}_N^e\times\mathbb{Z}_N^m=H^2(T^2,\mathbb{Z}_N^e\times\mathbb{Z}_N^m)~.
\end{equation}
Let us denote $H_\text{Gauss law}^{(0)}=H_\text{Gauss law}$, $H_\text{flux}^{(0)}=H_\text{flux}$.

We note that since $H^4(T^2,\mathbb{Z})$ is trivial, there is no higher Berry phase for this example.

\subsubsection{Coupling to electric particle}

We keep the same flux term, {\it i.e. }$\beta=0$, while modifying the Gauss law term into
\begin{equation}\label{eqn:Gausslaw}
    H_\text{Gauss law}^{(\alpha)}=-\sum_{(x,y)}{\cal X}_{(x-1,y),(x,y)}{\cal X}^\dag_{(x,y),(x+1,y)}{\cal X}_{(x,y-1),(x,y)}{\cal X}^\dag_{(x,y),(x,y+1)} \exp\left( \frac{2\pi i\alpha }{N}\psi^\dag(x,y)\psi(x,y)\right)+\text{h.c.}~.
\end{equation}
Such modification of the Gauss law term implies that at low energy, the electric charge equals to $\alpha$ multiple of  the fermion number modulo $N$.

For the Hamiltonian to commute with the Gauss law term, we modify the Hamiltonian with small Wilson line ${\cal Z}^\alpha=({\cal Z})^\alpha$:
\begin{align}
    H^{(\alpha,0)}=&\sum \psi^\dag(x,y)\left(\Gamma^1
    (m+c(\cos\lambda^z+\cos\lambda^w))+\Gamma^4 \sin\lambda^z+\Gamma^5\sin\lambda^w
    \right)\psi(x,y)\cr 
    &+\frac{c}{2}\sum \left(
    \psi(x+1,y)^\dag {\cal Z}^\alpha_{(x,y),(x+1,y)} 
    +\psi^\dag(x-1,y) {\cal Z}^{-\alpha}_{(x-1,y),(x,y)}
    \right.
    \cr
    &\qquad\qquad  \left.+
    \psi(x,y+1)^\dag {\cal Z}^\alpha_{(x,y),(x,y+1)} +\psi^\dag(x,y-1) {\cal Z}_{(x,y-1),(x,y)}^{-\alpha}
    \right)\Gamma^1 \psi(x,y)\cr
    &+\frac{1}{2i}\sum  \left(
    \psi(x+1,y)^\dag{\cal Z}_{(x,y),(x+1,y)}^\alpha -\psi^\dag(x-1,y){\cal Z}_{(x-1,y),(x,y)}^{-\alpha}
    \right)\Gamma^2 \psi(x,y)\cr
    &+\frac{1}{2i}\sum  \left(\psi(x,y+1)^\dag
    {\cal Z}_{(x,y),(x,y+1)}^\alpha
    -\psi^\dag(x,y-1)
    {\cal Z}_{(x,y-1),(x,y)}^{-\alpha}
    \right)\Gamma^3 \psi(x,y)~.
\end{align}

The family of gapped fermion theory coupled to $\mathbb{Z}_N$ topological order has Hamiltonian
\begin{align}
    &H_\text{new}^{(\alpha,0)}(\lambda^z,\lambda^w)=H^{(\alpha,0)}(\lambda^z,\lambda^w)+H^{(\alpha)}_\text{Gauss law}+H^{(0)}_\text{flux}~,\cr 
    &(\lambda^z,\lambda^w)\in M=T^2,\quad 
    \alpha\in H^2( M,\mathbb{Z}_N^e)=\mathbb{Z}_N^e
    ~.
\end{align}

\subsubsection{Coupling to magnetic particle}

We keep the same Gauss law term, {\it i.e.} $\alpha=0$, while modifying the flux term into
\begin{equation}\label{eqn:flux}
    H^{(\beta)}_\text{flux}=-\sum_{(x,y)}{\cal Z}_{(x,y),(x+1,y)} {\cal Z}_{(x,y+1),(x+1,y+1)} {\cal Z}_{(x,y+1),(x+1,y+1)}^\dag {\cal Z}_{(x,y),(x,y+1)}^\dag \exp\left( \frac{2\pi i\beta }{N}\psi^\dag(x,y)\psi(x,y)\right)+\text{h.c.}~,
\end{equation}
where we have chosen a framing to put the vertex that supports the charge of fermions at the lower left corner of the square that supports the $\mathbb{Z}_N$ flux.
Such modification of the flux law term implies that at low energy, the magnetic charge equals to $\beta$ multiple of  the fermion number modulo $N$.

For the Hamiltonian to commute with the flux term, we modify the Hamiltonian with small magnetic line ${\cal X}^\beta=({\cal X})^\alpha$:
\begin{align}
    H^{(0,\beta)}=&\sum \psi^\dag(x,y)\left(\Gamma^1
    (m+c(\cos\lambda^z+\cos\lambda^w))+\Gamma^4 \sin\lambda^z+\Gamma^5\sin\lambda^w
    \right)\psi(x,y)\cr 
    &+\frac{c}{2}\sum \left(
    \psi(x+1,y)^\dag {\cal X}^\beta_{(x+1,y),(x+1,y+1)} 
    +\psi^\dag(x-1,y) {\cal X}^{-\beta}_{(x,y),(x,y+1)}
    \right.
    \cr
    &\qquad\qquad  \left.+
    \psi(x,y+1)^\dag {\cal X}^\beta_{(x,y+1),(x+1,y+1)} 
    +\psi^\dag(x,y-1) {\cal X}_{(x,y),(x+1,y)}^{-\beta}
    \right)\Gamma^1 \psi(x,y)\cr
    &+\frac{1}{2i}\sum  \left(
    \psi(x+1,y)^\dag{\cal X}_{(x+1,y),(x+1,y+1)}^\beta -\psi^\dag(x-1,y){\cal X}_{(x,y),(x+1,y)}^{-\beta}
    \right)\Gamma^2 \psi(x,y)\cr
    &+\frac{1}{2i}\sum  \left(\psi(x,y+1)^\dag
    {\cal X}_{(x,y+1),(x+1,y+1)}^\beta
    -\psi^\dag(x,y-1)
    {\cal X}_{(x,y),(x+1,y)}^{-\beta}
    \right)\Gamma^3 \psi(x,y)~.\cr
\end{align}

The family of gapped fermion theory coupled to $\mathbb{Z}_N$ topological order has Hamiltonian
\begin{align}
    &H_\text{new}^{(0,\beta)}(\lambda^z,\lambda^w)=H^{(0,\beta)}(\lambda^z,\lambda^w)+H^{(0)}_\text{Gauss law}+H^{(\beta)}_\text{flux}~,\cr 
    &(\lambda^z,\lambda^w)\in M=T^2,\quad 
    \beta\in H^2( M,\mathbb{Z}_N^m)=\mathbb{Z}_N^m
    ~.
\end{align}

\subsubsection{Coupling to general dyonic particle}

Let us modify the Gauss law and flux term into $H_\text{Gauss law}^{(\alpha)}$ and $H_\text{flux}^{(\beta)}$ as above in (\ref{eqn:Gausslaw}) and (\ref{eqn:flux}) with general $(\alpha,\beta)$. For them to commute with the fermion Hamiltonian, the fermion Hamiltonian needs to be modified as
\begin{align}
    &H^{(\alpha,\beta)}\cr &=\sum \psi^\dag(x,y)\left(\Gamma^1
    (m+c(\cos\lambda^z+\cos\lambda^w))+\Gamma^4 \sin\lambda^z+\Gamma^5\sin\lambda^w
    \right)\psi(x,y)\cr 
    &+\frac{c}{2}\sum \left(
    \psi(x+1,y)^\dag {\cal Z}^{\alpha}_{(x,y),(x+1,y)} {\cal X}^\beta_{(x+1,y),(x+1,y+1)} 
    +\psi^\dag(x-1,y){\cal Z}^{-\alpha}_{(x-1,y),(x,y)} {\cal X}^{-\beta}_{(x,y),(x,y+1)}
    \right.
    \cr
    &\qquad\qquad  \left.+
    \psi(x,y+1)^\dag {\cal Z}^{\alpha}_{(x,y),(x,y+1)} {\cal X}^\beta_{(x,y+1),(x+1,y+1)} 
    +\psi^\dag(x,y-1){\cal Z}^{-\alpha}_{(x,y-1),(x,y)} {\cal X}_{(x,y),(x+1,y)}^{-\beta}
    \right)\Gamma^1 \psi(x,y)\cr
    &+\frac{1}{2i}\sum  \left(
    \psi(x+1,y)^\dag
    {\cal Z}^{\alpha}_{(x,y),(x+1,y)}
    {\cal X}_{(x+1,y),(x+1,y+1)}^\beta -\psi^\dag(x-1,y)
    {\cal Z}^{-\alpha}_{(x-1,y),(x,y)}
    {\cal X}_{(x,y),(x+1,y)}^{-\beta}
    \right)\Gamma^2 \psi(x,y)\cr
    &+\frac{1}{2i}\sum  \left(\psi(x,y+1)^\dag
    {\cal Z}^{\alpha}_{(x,y),(x,y+1)}
    {\cal X}_{(x,y+1),(x+1,y+1)}^\beta
    -\psi^\dag(x,y-1)
    {\cal Z}^{-\alpha}_{(x,y-1),(x,y)}
    {\cal X}_{(x,y),(x+1,y)}^{-\beta}
    \right)\Gamma^3 \psi(x,y)~,\cr
\end{align}
where we choose an ordering of ${\cal Z},{\cal X}$ in each term to make the Hamiltonian hermitian. For any ordering, the modified fermion Hamiltonian still commutes with the modified Gauss law and modified flux terms.

The family of gapped fermion theory coupled to $\mathbb{Z}_N$ topological order has Hamiltonian
\begin{align}
    &H_\text{new}^{(\alpha,\beta)}(\lambda^z,\lambda^w)=H^{(\alpha,\beta)}(\lambda^z,\lambda^w)+H^{(\alpha)}_\text{Gauss law}+H^{(\beta)}_\text{flux}~,\cr 
    &(\lambda^z,\lambda^w)\in M=T^2,\quad 
    (\alpha,\beta)\in H^2( M,\mathbb{Z}_N^e\times\mathbb{Z}_N^m)=\mathbb{Z}_N^e\times\mathbb{Z}_N^m
    ~.
\end{align}

\subsubsection{Fractional Berry phase: two copies of Chern insulator coupled to toric code}
\label{sec:exampleT4fracBerryphase}

The 2+1d theory with parameter space $M=T^2$ does not have non-trivial Berry phase due to vanishing $H^4(M,\mathbb{Z})$. On the other hand, we can consider two copies of the Chern insulator coupled to the same $\mathbb{Z}_N$ toric code, $M=T^4$, then it is possible to have non-trivial Berry phase. 
For the two copies of Chern insulator coupled to the toric code by $(\alpha,\beta;\alpha',\beta')$,
the Berry phase is in fact fractional, given by the braiding of the dyons of electric/magnetic charges $(\alpha,\beta)$ and $(\alpha',\beta')$. Denote the coordinate on the two $T^2$ by $\lambda^z,\lambda^w$ and $\lambda'^z,\lambda'^w$, respectively, the Berry curvature is given by\footnote{
One can also compute the higher Berry curvature from the formula in \cite{Kapustin:2020eby}, which we will leave for future work.} 
\begin{align}
&{2\pi \over (2\pi)^4 N} \left(\alpha d\lambda^zd\lambda^w+\alpha' d\lambda'^zd\lambda'^w\right)
    \left(\beta d\lambda^zd\lambda^w+\beta' d\lambda'^zd\lambda'^w\right)\cr
    &=\frac{2\pi}{N(2\pi)^4}\left(\alpha\beta'+\alpha'\beta\right)d\lambda^zd\lambda^wd\lambda'^zd\lambda'^w~.
\end{align}

\section{Phase transitions protected by family of topological phases}
\label{sec:phase transition}

Suppose we have a family of systems labelled by parameters that live in space ${\tilde M}$. For generic value of the parameter, the systems are typically gapped, where the energy gap protects the system against small perturbations. We can divide ${\tilde M}$ into region $M$ that label family of gapped systems, and region $M'$ with gapless phase transitions, surrounded by the gapped phases. We would like to understand how to use the topological property of the gapped phases in $M$ to learn the presence of these gapless phase transitions $M'$.

Suppose we have non-trivial family of gapped systems in region $M$ of parameter space.
If such family becomes contractible in the enlarged parameter space ${\tilde M}$, then ${\tilde M}$ must contain gapless phase transitions $M'$.

If we vary the parameter over $n$-dimensional cycle in spacetime, with value tracing out some $k$-dimensional cycle in the parameter space $M$, the $n$-dimensional cycle intersects with a non-trivial topological defect of codimension $n$. (The codimension-$(n+1)$ excitations created by the topological defect are contained in the $n$-dimensional locus). 
We will say the defect is trapped by the cycle. If the parameter space can be enlarged to $\widetilde M=M\cup M'$ such that the $k$-cycle becomes contractible, then the gapped system cannot extend smoothly over $\widetilde M$, but there must be a phase transition. Moreover, at the phase transition, the topological defect trapped by the cycle can be screened.

In the special case that the gapped systems are invertible, the phase transitions on ${\widetilde M}$ are called diabolical loci and they are studied in \cite{Hsin:2020cgg}. It is described as an ``anomaly in the space of coupling" in \cite{Cordova:2019jnf,Cordova:2019uob}.
Here, we generalize the discussion to the case where the phases are gapped but can be non-invertible.
In such case, describing the family (and thus the phase transitions) using the Berry phase is not sufficient, since some Berry phase can become trivial. In the terminology of \cite{Cordova:2019jnf,Cordova:2019uob}, such phase transitions may not have an anomaly in the space of coupling.
Nevertheless, such phase transitions without an anomaly in the space of coupling can still be protected by the non-trivial family of gapped systems.

\subsection{Example: $\mathbb{Z}_2$ gauge field coupled to  free Dirac fermion in 1+1d}

Let us consider $\mathbb{Z}_2$ gauge theory in 1+1d coupled to a Dirac fermion of charge one by the $U(1)_V$ symmetry, and we deform the theory by a complex mass term that breaks $U(1)_A$ explicitly
\begin{equation}
    m\bar\psi \left(\cos\theta+i\gamma^{01}\sin\theta\right)\psi~,
\end{equation}
where $\theta$ shifts under a $U(1)_A$ transformation.
For $m\neq 0$ we have a family of gapped systems over $M=S^1$. If we integrate out the fermions for $m\neq 0$, this produces the effective action \cite{Abanov:1999qz}
\begin{equation}
    \frac{1}{2\pi}\int a d\theta~,
\end{equation}
where $a$ is the $\mathbb{Z}_2$ gauge field. In other words, if we vary the $\theta$ parameter by $2\pi$ around a non-contractible cycle, we will trap a $\mathbb{Z}_2$ Wilson line.

If we extend the parameter space to include $m=0$, $\widetilde M=\mathbb{R}^2$, then the $S^1$ cycle becomes contractible in $\widetilde M$, and we conclude there is a phase transition on $\widetilde M$. The phase transition at $m=0$ is described by $\mathbb{Z}_2$ gauge theory coupled to a massless Dirac fermion, the theory is free, and the Wilson line is screened by the charge-one massless Dirac fermion.

\subsection{Example: toric code topological order with parameter $S^1$ in 2+1d}

Examples of such phase transition on ${\widetilde M}$ for non-trivial family of gapped system on $  M \cong S^1$ is described by the model in \cite{Kamfor:2010ie, PhysRevB.91.134419} for ${\widetilde M}=D^2$, which is a deformation of the model in \cite{Kitaev:2005hzj}. The nontrivial-ness of the family is characterized by $\pi_1(M)=\mathbb{Z}$ as discussed in \cite{Aasen:2022cdu}.
The family of gapped systems has toric code topological order, and going around the parameter space $S^1$ the system is transformed by the $e,m$ exchange symmetry.
The non-trivial family of gapped systems protects the gapless point inside ${\widetilde M}\cong D^2$, which is described by the $(\mathbb{Z}_2)_{0}$ gauge theory coupled to massless Dirac fermion by gauging the fermion parity symmetry.

\subsection{Example: Abelian Higgs model in 2+1d}

Let us give an example where the family has non-trivial element in $H^2(M,{\cal A})$. Consider the example of Abelian Higgs model coupled to $\mathbb{Z}_N$ gauge theory, with parameter space $M=S^2$. The parameter is given by the $SO(3)$-triplet mass of the Higgs scalars, and the family can have non-trivial $H^2(S^2,\mathbb{Z}_N)$. 
We can extend the parameter space to be ${\widetilde M}=B^3$, whose boundary is $S^2$, and $f$ is the boundary map. Then we conclude there must be a phase transition inside $B^3$, which is the deconfined quantum criticality.
We note that the Abelian anyon can have non-trivial statistics, and we cannot get rid of the Abelian anyon by condensation.

For instance, let us consider the Abelian Higgs model in 2+1d coupled to $U(1)_k$ Chern-Simons gauge field. Then the phase transition at small mass is protected not by Berry phase (there is none), but by the nontrivial family of gapped systems with $U(1)_k$ topological order.

\section{Bulk-boundary correspondence}
\label{seC:bulk-boundary}

In this section, we will investigate the bulk-boundary correspondence for family of gapped systems with the same topological order. 
In the case that the underlying bulk topological order is invertible, the boundary is well-defined, and we can ask if the boundary can be trivially gapped for all parameters. The answer is negative: this is the bulk-boundary correspondence between the bulk Berry phase \cite{Hsin:2020cgg}, and the phase transition on the boundary can be detected by an anomaly in the space of coupling \cite{Cordova:2019jnf}.

More generally, the family of gapped systems in the bulk can have non-trivial topological order, then the boundary is a relative theory \cite{Freed:2012bs}.\footnote{We can obtain a well-defined ``absolute" boundary theory from the relative theory by choosing a polarization, which is equivalent to a choice of topological boundary condition (or more generally, topological domain wall to an invertible phase). The absolute theory corresponding to the the relative theory is obtained by putting the bulk theory on an interval with the topological boundary condition on one end and the relative theory on the other end, and shrinking the interval to collide the relative theory with the topological boundary condition \cite{Gukov:2020btk}.}
We will consider the case that the bulk topological order admits a topological boundary condition, since if the boundary is gapless the boundary can have phase transitions by adding relevant deformations.\footnote{It would also be interesting if the bulk-boundary correspondence can rule out isolated gapless theories without relevant deformations.}
We ask the following question: if the bulk topological order comes from a non-trivial family of gapped systems, can we have a family of gapped boundary systems with the same underlying boundary topological order given by the topological boundary condition, without closing the energy gap at any value of the parameter.

\subsection{Review of bulk-boundary correspondence for invertible bulk topological order}

Let us begin by reviewing the case when the bulk is a family of invertible phase, as discussed in Section VI of \cite{Hsin:2020cgg}. Let us consider non-trivial family, such that when we promote the parameter to be spacetime-dependent $\lambda:X\rightarrow M$ on spacetime manifold $X$, it is topologically non-trivial, and the partition has non-trivial winding. 
In other words, the partition function transforms non-trivially when we transit between different coordinate patches on $M$. 
For instance, if the parameter space is $M=S^1$, parameterized by $\theta\in S^1$, the partition function has non-trivial winding. We can undo such winding using a boundary (or interface). Then the partition function in the presence of such boundary must vanish at some parameter, and the gap must close on the boundary, at the parameter.
Explicitly, for the transition function $\theta\rightarrow \theta+2\pi n$ on $M$ with integer $n$, the partition function transforms by a phase $e^{i\phi(n)}$; thus demanding the partition function to be a function of the parameter space, we find that the partition function must vanish at some parameter $\theta=\theta^*$. (Later on, we will generalize the discussion to the case that the partition function transforms under a transition function on $M$ by the insertion of a topological defect instead of the identity operator multiplied with a phase).

For general $M$, we can interpret such ``anomaly inflow" using the transition function on coordinate patches on the parameter space.
The non-trivial family of invertible phases in the bulk implies that on the boundary, we cannot define the partition function as a well-defined function on the parameter space unless the partition function vanishes at some parameter, since when we move across different coordinate patches, the partition function has non-trivial transition function. If we remove these parameters where the boundary partition function vanishes, then the transition function is the Berry phase on the boundary. The bulk-boundary correspondence is the analogue of clutching construction of vector bundles that infers the bulk Berry phase using the transition functions on the boundary.

\subsection{Bulk-boundary correspondence for general bulk topological order}

Consider family of gapped systems in the bulk with the same non-trivial topological order. We assume the topological order admits a topological boundary condition. And to derive a contradiction, let us assume we have family of gapped systems on the boundary over the same parameter space, with the same topological order given by the topological boundary condition without closing the gap.

Let us consider the family of gapped systems protected by $\pi_n(M)$. In other words, if we vary the parameter over some non-contractible $n$-dimensional cycle in spacetime with value tracing out some non-contractible $k$-dimensional cycle in the parameter space for $k\leq n$, the locus traps a non-trivial topological defect of codimension $n$.
On the boundary, we can do the same thing. However, some cycles on the boundary become contractible due to the bulk. We conclude that the corresponding topological defect must condense on the boundary; if this is not possible on the choice of topological boundary condition we start with, then we conclude the boundary must have a phase transition.
We can also repeat the argument in \cite{Hsin:2020cgg}, where the winding of the parameter produces a topological defect instead of a phase in the partition function.
We can unwind the parameter in the bulk using the boundary, and thus the corresponding topological defect must become trivial on the boundary at some parameter. If the topological defect does not condense on the choice of topological boundary condition, then the boundary must have a phase transition at some parameter driven by the condensation of such topological defect.
Thus both arguments imply that the gap must close on the boundary at some parameter for the topological boundary condition where the topological defect does not condense.

\subsection{Example: free fermion in 1+1d}

Let us consider free Dirac fermion in 1+1d coupled to $\mathbb{Z}_2$ gauge field by $\mathbb{Z}_2\subset U(1)_V$ symmetry (we can also deform the theory by fermion bilinear with small coefficient to break the $U(1)_V$ to $\mathbb{Z}_2$), and we turn on the complex mass parameter $me^{i\theta}$ with $\theta$ shifts under a $U(1)_A$ rotation.
For $m\neq 0$, we can integrate out the massive fermions, and for spacetime dependent background field $\theta$ this gives  the effective action
\begin{equation}
    \frac{1}{2\pi}\int a d\theta~,
\end{equation}
where $a$ is the $\mathbb{Z}_2$ gauge field. The theory at low energy is the $\mathbb{Z}_2$ gauge theory in 1+1d, and it has two vacua on a circle.

Let us consider the interface across which $\theta$ continuously interpolates from $\theta=\theta_0$ to $\theta=\theta_0+2\pi$ in the direction normal to the interface. 
Suppose the interface is described by a family of gapped system with $\mathbb{Z}_2$ gauge field, then it can be described by Chern-Simons term
\begin{equation}
    k(\theta_0)\int a~,
\end{equation}
where the function $k(\theta_0)$ is an integer, which can have discontinuous jump when we vary $\theta_0$ at phase transitions. As we vary $\theta_0$ by $2\pi$, the family of gapped system in the bulk implies that the interface has extra $\int a$, and thus for $k(\theta_0)=k(\theta_0+2\pi)$ to be a well-defined function of $\theta_0\in S^1$, the Wilson line $\int a$ must become trivial at some $\theta_0=\theta_0'$. 

\subsection{Example: gauge theory theta angle in 3+1d}

Consider bulk 4+1d two-form $\mathbb{Z}_N$ gauge theory. The theory has electric and magnetic strings, with charge $(q_e,q_m)\in\mathbb{Z}_N\times\mathbb{Z}_N$.
As discussed in \cite{Gaiotto:2014kfa,Chen:2021xuc}, the theory has $SL(2,\mathbb{Z}_N)$ 0-form symmetry, generated by the transformations $S:(q_e,q_m)\rightarrow (q_m,-q_e)$ and $T:(q_e,q_m)\rightarrow (q_e+q_m,q_m)$. In particular, the generator of the $T$ transformation is the domain wall \cite{Chen:2021xuc}
\begin{equation}\label{eqn:4+1dT}
    \frac{2\pi(N-1)}{2N}\oint_{M_4} {\cal P}(b_2)~,
\end{equation}
where $M_4$ is a four-dimensional submanifold that supports the domain wall, $b_2$ is the $\mathbb{Z}_N$ two-form gauge field, and ${\cal P}$ is the Pontryagin square operation. Let us consider a family of gapped system in 4+1d with $\mathbb{Z}_N$ two-form gauge theory topological order, parameterized by $\theta\in S^1$, such that when we vary $\theta$ by $2\pi$ along a circle, there is topological domain wall (\ref{eqn:4+1dT}) intersecting the circle. If we promote the theta angle to be spacetime-dependent, this is described by the coupling
\begin{equation}\label{eqn:ZNbulk}
    \frac{2\pi(N-1)}{2N}\int \frac{d\theta}{2\pi} {\cal P}(b_2)~.
\end{equation}

\subsubsection{Prediction of phase transition with non-invertible symmetry}

A boundary theory of the two-form gauge theory is $SU(N)$ or $PSU(N)$ Yang-Mills theory in 3+1d, where $\theta$ is the theta angle that couples to the instanton number, and the theory couples to the bulk by the one-form symmetry. We can choose a polarization to obtain $SU(N)$ Yang-Mills theory, where the two-form gauge field obeys Dirichlet boundary condition, or $PSU(N)$ Yang-Mills theory, where the two-form gauge field obeys Neumann boundary condition.

Consider $SU(N)$ Yang-Mills theory, it has $\mathbb{Z}_N$ center one-form global symmetry with a Berry phase for the parameter space $S^1$ for the theta angle, described by the bulk effective action (\ref{eqn:ZNbulk}).
The Berry phase implies that there must be a phase transition on the boundary at some theta $\theta_\star$. Arguments using mixed anomaly between time-reversal symmetry and one-form symmetry as well as soft supersymmetry breaking suggest a first order phase transition at $\theta_\star=\pi$ with spontaneously broken time-reversal symmetry, see {\it e.g.} \cite{Gaiotto:2017yup}.\footnote{
We remark that Standard Model can have center one-form symmetry $\mathbb{Z}_3,\mathbb{Z}_2,$ trivial, or $\mathbb{Z}_6$, depending on the choices of gauge group, which can be $\left(SU(3)\times SU(2)\times U(1)\right)/C$ for $C=\mathbb{Z}_2,\mathbb{Z}_3,\mathbb{Z}_6$ or trivial \cite{Tong:2017oea}; on the other hand, the effective theta angle (including the contribution from the massive fermions) is small $\theta\sim 10^{-11}$.
}

Now, let us gauge the one-form symmetry: this gives $PSU(N)$ Yang-Mills theory. Equivalently, we consider Neumann boundary condition for the two-form gauge field.
For such a boundary condition, the operator $\oint b_2$ is non-trivial on the boundary, and the topological defect (\ref{eqn:4+1dT}) is non-trivial on the boundary.
Thus there must be a phase transition on the boundary, where the 
topological defect (\ref{eqn:4+1dT}) is screened. Since the theory is believed to have a mass gap, the screening is implemented by a three-dimensional topological defect as topological boundary condition for (\ref{eqn:4+1dT}). Moreover, from the topological defect (\ref{eqn:4+1dT}), we can compute the fusion rule of the three-dimensional topological defect at the phase transition. Using the methods in \cite{Choi:2022zal,Kaidi:2021xfk,Barkeshli:2022edm}, we find that the three-dimensional topological defect, denoted by $U(M_3)$ on three-dimensional submanifold $M_3$, obey non-invertible fusion rule\footnote{
For instance, using the method of \cite{Barkeshli:2022edm}, let us express the domain wall (\ref{eqn:4+1dT}) by embedding the $\mathbb{Z}_N$ two-form gauge field in a continuous $U(1)$ two-form gauge field $b$ with $\oint b\in (2\pi/N)\mathbb{Z}$ as
\begin{equation}
    \frac{N(N-1)}{4\pi}\int bb~.
\end{equation}
Under a one-form gauge transformation $b\rightarrow b+d\lambda$ where $\lambda$ is a $U(1)$ gauge field that embeds the $\mathbb{Z}_N$ one-form parameter, the above term produces the inflow boundary term
\begin{equation}
\int_{M_3}\left(    \frac{N(N-1)}{4\pi} \lambda d\lambda+\frac{N(N-1)}{2\pi}b\lambda\right)~.
\end{equation}
Since $\oint\lambda\in (2\pi/N)\mathbb{Z}$, the parameter $\lambda$ is related to $\gamma$ by $N\lambda/2\pi=\text{PD}(\gamma)$, and the inflow gives the fusion rule (\ref{eqn:ZNYMfusion}).
}
\begin{equation}\label{eqn:ZNYMfusion}
    U(M_3)\times \overline{U(M_3)}= \frac{1}{{\cal N}} \sum_{\gamma\in H_2(M_3,\mathbb{Z}_N)}\eta(\gamma)(-1)^{(N-1)\int_{M_3}\text{PD}(\gamma)\cup d \text{PD}(\gamma)/N}~,
\end{equation}
where $\eta(\gamma)=e^{(2\pi i/N)\oint_\gamma b_2}$, PD denotes Poincar\'e dual, and ${\cal N}=|H^0(M_3,\mathbb{Z}_N)|$ is a normalization factor.
In other words, the family of gapped system in the bulk predicts the theory has a phase transition, and moreover it predicts the phase transition has non-invertible symmetry generated by topological domain wall $U(M_3)$. This is consistent with the result in \cite{Choi:2022zal,Kaidi:2021xfk}.

 \subsubsection{Axion Yang-Mills theory and non-invertible higher-group symmetry}

Let us study the consequence of the bulk-boundary correspondence by promoting the theta angle to be dynamical axion field $\theta=La/f_a$ for axion decay constant $f_a$, and some integer $L$.
 We remark that the axion field is originally introduced to solve the strong CP problem in Standard Model \cite{PhysRevLett.38.1440}, such that the expectation value of the axion from a potential sets the effective theta angle to zero, see {\it e.g.} \cite{Blinov:2022tfy} for a recent review.
Consider $PSU(N)$ gauge theory coupled to the axion field $a$. Let us consider pure Yang-Mills theory, while charged fermions can be included in a straightforward way.
The theory has the non-invertible topological domain wall $U(M_3)$ that generates the discrete shift symmetry $a\rightarrow a+2\pi f_a/L$. Across the domain wall, the two sides differ by a  non-trivial discrete theta angle \cite{Aharony:2013hda,Hsin:2020nts}, and the spectrum of line operators change \cite{Aharony:2013hda}. 

As another example, consider $Sc(N)$ Yang-Mills theory with $N=0$ mod 4 coupled to the axion field $a$, which is related to $Spin(N)$ axion Yang-Mills theory by gauging the $\mathbb{Z}_2$ center one-form symmetry that acts on the Wilson lines in the representations with odd number of boxes in the young tableaux.\footnote{
For $N=0$ mod 4, the center of $Spin(N)$ has three $\mathbb{Z}_2$ subgroups, and the three quotients $Spin(N)/\mathbb{Z}_2$ for these three $\mathbb{Z}_2$ subgroups give $SO(N),Sc(N)$ and $Ss(N)$.
} 
Across the domain wall where $a\rightarrow a+2\pi f_a/L$, the theory changes by
\begin{equation}
\frac{2\pi N}{16}\int {\cal P}(w_2^{Sc})+
    \pi\int\left( w_2^{Sc}\cup B_2+B_2\cup B_2\right)~,
\end{equation}
where $w_2^{Sc}$ is the obstruction to lifting the $Sc$ bundle to an $Spin$ bundle, and it is the $\mathbb{Z}_2$ magnetic flux in the $Sc(N)$ gauge theory, and $B_2$ is the background for the center one-form symmetry in $Sc(N)$ gauge theory that acts on Wilson lines in the spinorial representations. The first term can be compensated by a TQFT on the domain wall, which makes the domain wall non-invertible.
The coupling to $w_2^{Sc}$ implies that across the domain wall, the fractionalization class changes \cite{Benini:2018reh}. In particular, if we intersects the generator for the center one-form symmetry with the domain wall that generates the shift in $a$ at a one-dimensional intersection locus, then the term implies the intersection emits the generator for the magnetic one-form symmetry $\oint w_2^{Sc}$. 
Moreover, the last term implies that for the configuration of generator of the generator of the center one-form symmetry with non-trivial self-intersection, it can be removed by shifting $a$, {\it i.e.} there is $\int da$ emitted from the intersection point. The property that intersection of symmetry defects produces additional symmetry defect means that these symmetries mix into the analogue of three-group, but it involves non-invertible symmetry generated by the shift of $a$.

We remark that similar consideration can also show that the Maxwell $U(1)$ gauge theory with axion has non-invertible symmetry, reproducing the result in \cite{Cordova:2022ieu}, as well as higher-group symmetry \cite{Brennan:2020ehu}.

\section*{Acknowledgments}
The authors thank M. Cheng, C.-M. Jian, A. Kapustin, A. Kitaev, D. Ranard and X.-G. Wen for discussions. The authors thank M. Cheng and C. C\'ordova for comments on a draft.
The second author thanks A. Kitaev for discussion on potential issues in a mathematically rigorous definition of the space $\mathcal{M}_\B$ and physical interpretations of $\pi_3(B\uuPic\B)$.
Z.W. is partially supported by NSF grants
FRG-1664351, CCF 2006463, and ARO MURI contract W911NF-20-1-0082. P.-S.H. is supported by Simons Collaboration on Global Categorical Symmetries. 
The authors thank the American Institute of Mathematics for hosting the workshop ``higher categories and topological order” where this work was partially performed.

\bibliographystyle{utphys}
\bibliography{main}

\end{document}